\documentclass[12pt,a4paper]{iopart}
\usepackage{iopams}
\usepackage{graphicx}
\usepackage[usenames]{color}
\usepackage[normalem]{ulem}
\newcommand{\cpc}{\emph{Comp. Phys. Commun.\ }}

\begin{document}

\title{Three-dimensional vortex structures in a rotating dipolar Bose-Einstein condensate}

\author{Ramavarmaraja Kishor Kumar$^{1}$, Thangarasu Sriraman$^{2}$, Henrique Fabrelli$^{1}$, 
 Paulsamy Muruganandam$^{2}$, and Arnaldo Gammal$^{1}$}
\address{$^1$Instituto de F\'{i}sica, Universidade de S\~{a}o Paulo, 05508-090 S\~{a}o Paulo, Brazil \\
$^2$ School of Physics, Bharathidasan University, Palkalaiperur Campus,
Tiruchirappalli 620024, Tamilnadu, India}

\begin{abstract} We study three-dimensional vortex lattice structures in purely dipolar 
Bose-Einstein condensate (BEC). By using the mean-field approximation, we obtain a stability diagram 
for the vortex states in purely dipolar BECs as a function of harmonic trap aspect ratio 
($\lambda$) and dipole-dipole interaction strength ($D$) under rotation. Rotating the condensate 
within the unstable region leads to collapse while in the stable region furnishes stable vortex 
lattices of dipolar BECs. We analyse stable vortex lattice structures by solving the three-dimensional 
time-dependent Gross-Pitaevskii equation in imaginary time. Further, the stability of vortex states 
is examined by evolution in real-time. We also investigate the distribution of vortices in a fully 
anisotropic trap by increasing eccentricity of the external trapping potential. We observe the breaking  
 up of the condensate in two parts with an equal number of vortices on each when the trap is 
sufficiently weak, and the rotation frequency is high.

\end{abstract}

\maketitle
\section{Introduction}

Vortices in Bose-Einstein condensates of alkali atoms first observed in the laboratory 
in 1999~\cite{Matthews1999}. Since then numerous experimental and theoretical studies along 
this direction were performed~\cite{review}. For 
instance, these studies include bending of vortex lines in a cigar shaped trap, array of orderly 
aligned lattices in the quantum-Hall regime, Tkachenko oscillations in the lowest Landau level and so 
on~\cite{vortex-pending,Tkachen}. The rotational properties of BECs of alkali atoms are reviewed in 
Refs~\cite{review,Minguzzi2004}. Several techniques such as rotating the magnetic trap, laser 
stirring, decay of solitons, imprinting vortices using topological phases, superimposing an 
oscillating excitation to the trapping potential, and applying artificial magnetic 
fields~\cite{review,Exp,Gauge} were adopted in BEC experiments to nucleate the vortices.

Early experiments and theoretical studies on vortices in BEC mostly focused on alkali Bose gases with local 
and isotropic interaction. In particular, there are many numerical studies proposed to analyse the 
stationary state of rotating alkali BECs~\cite{Aft-Bak-Jen}. However, the experimental realization of Bose-Einstein 
condensation in chromium ($^{52}$Cr)~\cite{Gries-cr}, dysprosium ($^{164}$Dy)~\cite{Lu-dy}, followed by erbium 
($^{168}$Er)~\cite{Aika-er} has enlightened new directions in understanding the properties of BEC in dipolar 
quantum gases. These recent works has revealed new phenomena due to the peculiar competition 
between isotropic short-range contact interaction and anisotropic long-range dipole-dipole interaction (DDI). The 
most significant features of the dipolar BECs are the emergence of biconcave shaped states, the dependence 
of the stability on the trap geometry, the roton-like dip in the dispersion relation for excitation waves, and the 
$d$-wave mode of the collapse~\cite{Baranov2008,Lahaye2009, Young2012,Andreev2014}. The rotational properties of 
dipolar Bose gases have also been studied theoretically in the mean-field 
regime~\cite{Yi2006,Wilson2009,Abad2009,Van2007,Malet2011,Kishor2012}. These studies revealed that the rotational 
properties of dipolar BECs are strongly influenced by the harmonic trap aspect ratio, DDI 
strength, contact interaction (CI) strength and relative strengths between DDI and CI~\cite{Malet2011,Kishor2012}. 
Square and triangular vortex lattice structures have been predicted in dipolar BECs loaded in optical lattice 
potentials~\cite{Kishor2014}. The stability of dipolar BEC is strongly dependent on the trap geometry and dipolar 
interaction strength. So far the regime for the stability has been analysed only for the non-vortex states of 
dipolar BECs without rotation~\cite{Ronen2007,Dutta2007}. Most of the previous studies on vortices in dipolar BECs 
are based on two-dimensional models only~\cite{Yi2006,Wilson2009,Abad2009,Van2007,Malet2011}. However, it will be 
more realistic to investigate the stationary vortex structures in full three-dimensions as they are readily 
comparable with experiments.

In this paper, we study the vortex lattice structures of purely dipolar Bose-Einstein condensate by considering the 
full three-dimensional Gross-Pitaevskii equation. In particular, we analyse the stability regime for vortex state in purely 
dipolar BECs with respect to both trap aspect ratio and DDI strength. The stability of the vortices is 
confirmed by numerically evolving the vortex states in real-time. We show stationary vortex lattice structures for 
different trap aspect ratios within the stable regime. Further, we study the collapse dynamics of biconcave shaped 
condensate during rotation. We calculate the chemical potential, rms (root-mean-square) radius and angular 
momentum of condensate as a function of rotation frequency. We estimate the number of vortices 
using Thomas-Fermi approximation and compare them with that obtained through numerical simulations. 
Finally, we notice the breaking of the condensate when the rotating dipolar BEC is in the fully anisotropic trap. 


The paper is organized as follows. In Sec.~\ref{sec:frame}, we present the mean-field equation to 
study a dipolar BEC in a rotating trapping potential. In Sec.~\ref{sec:stab}, we show the stability 
diagram for states without and with vortices in purely dipolar BECs as a function of harmonic 
trap aspect ratio and DDI strength. In Sec.~\ref{sec:vortex}, we study the 
stationary vortex structures observed numerically as well Thomas-Fermi calculation on some physical 
parameters during rotation of the condensate. Then we investigate the spatial distribution of 
vortices in a fully anisotropic trap in Sec.~\ref{sec:ani}. Finally, in Sec.~\ref{sec:con}, we 
present a brief summary and conclusions.


\section{The mean-field Gross-Pitaevskii equation in rotating frame}
\label{sec:frame}
The pattern of vortices in the  BEC can be studied using the mean field Gross-Pitaevskii (GP) equation~\cite{review, Bao2006}. At ultra-low temperatures the properties of a dipolar Bose-Einstein condensate of $N$ atoms, can be described by the mean-field GP equation in rotating frame with the nonlocal nonlinearity  of the form~\cite{Yi2006,Van2007}:
\begin{eqnarray}
\mathrm{i}\hbar\frac{\partial \phi({\mathbf r},t)}{\partial t}  & =   \left[-\frac{\hbar^2}{2m}\nabla^2+V_{{trap}}({\mathbf r}) 
+ \frac{4\pi\hbar^2a N}{m}  \vert \phi({\mathbf r},t)\vert^2 \right. \nonumber \\
 &  \left.  -\Omega L_z  + N \int U_{\mathrm{dd}}({\mathbf  r}-{\mathbf r}') \vert\phi({\mathbf r}',t)\vert^2 d{\mathbf r}' 
\right]\phi({\mathbf r},t),
\label{eqn:dgpe}
\end{eqnarray}
where $\int d{\bf r}\vert\phi({\mathbf r},t)\vert^2=1.$ The trapping potential, $V_{{trap}}$ is assumed to be fully  
asymmetric of the form
\begin{eqnarray}
V_{{trap}}({\mathbf r}) = \frac{1}{2} m \left(\omega_x^2 x^2+\omega_y^2 y^2+ \omega_z^2 z^2 \right) \nonumber
\end{eqnarray}
where $\omega_x, \omega_y $ and  $\omega_z$ are the trap frequencies, $m$ is the atom mass, and $a$ is the atomic scattering length.  $L_z= - \mathrm{i} \hbar (x\partial_y-y \partial_x)$ corresponds to the $z$-component of the angular momentum due to the rotation of the dipolar BEC about $z$-axis with rotation frequency $\Omega$. 

For magnetic dipoles, the dipolar interaction is given by \cite{tuning-ddi,Goral2002} 
\begin{eqnarray}
U_{\mathrm{dd}}({\bf R}) = \frac{\mu_0 \bar \mu^2}{4\pi}\frac{1-3\cos^2 \theta}{ \vert  {\bf R} \vert  ^3} 
\left( \frac{3 \cos^2{\varphi}-1}{2}\right),
\end{eqnarray}
where ${\bf R= r -r'}$ determines the relative position of dipoles, $\theta$ is the angle between ${\bf R}$ and the direction of polarization, $\mu_0$ is the permeability of free space, and $\bar \mu$ is the dipole moment of the atom. $\varphi$ is the angle between the orientation of dipoles and $z$-axis. We consider the polarization of magnetic dipoles along the direction of $z$-axis as long as $\varphi$ = 0. 

To compare the contact and dipolar interactions, it is convenient to introduce the length scale $a_{\mathrm{dd}}\equiv \mu_0 \bar \mu^2  m/(12\pi \hbar^2)$ \cite{Koch2008}. $^{52}$Cr has a magnetic dipole moment of $\bar \mu = 6\mu_B$ ($\mu_B$ is the Bohr magneton) so that $a_\mathrm{dd} \simeq 16a_0$, where $a_0$ is the Bohr radius. The dipole-dipole interaction strength is expressed as $D=3 N a_{\mathrm{dd}}$.

Convenient dimensionless parameters can be defined in terms of a reference frequency $\bar \omega $ and the corresponding oscillator length $l=\sqrt{\hbar/(m\bar \omega)}$. Using dimensionless variables $\mathbf{r}' = {\bf r}/l, \mathbf{R}'= {\bf R}/l, a' = a/l, a_{\mathrm{dd}}'= a_{\mathrm{dd}}/l$, $t' = t\bar \omega$, $x' = x/l$, $y' = y/l$, $z' = z/l$, $\Omega'=\Omega/\bar \omega$,  $\phi' = l^{3/2}\phi$, equation (\ref{eqn:dgpe}) can be rewritten (after dropping the primes from all the variables) as
\begin{eqnarray}\label{gpe3d}
\mathrm{i} \frac{\partial \phi( {\mathbf {r}},{t})}{\partial t} = \left[-\frac{1}{2}\nabla^2+ \frac{1}{2}\left({\gamma^2 x^2+\nu^2  y^2}+\lambda^{2} z^2\right) + 4 \pi a N \vert {\phi} \vert  ^2 \right. \nonumber \\
 \left. - \Omega L_z  + 3 N a_\mathrm{dd} \int  V_{\mathrm{dd}}^{3D}({\bf { R}}) \vert \phi( {\mathbf { r}}', t) \vert  ^2 d
{\mathbf { r}}' \right]  \phi({\mathbf { r}},{ t}),
\end{eqnarray}
with
\begin{eqnarray}\label{vdd}
V_{\mathrm{dd}}^{3D}({\bf{R}})=\frac{1-3\cos^2\theta}{\vert{\bf{R}}\vert^3}\left(\frac{3\cos^2{\varphi}-1}{2}\right) \, ,
\end{eqnarray}
$\gamma= \omega_x/\bar \omega, \nu=\omega_y/\bar \omega, \lambda =\omega_z/\bar \omega$. We consider the cylindrically symmetric harmonic trap with $\gamma=\nu$ with $\omega_x=\omega_y=\omega_\rho$ and we use the reference  frequency $\bar \omega$ as  $\omega_\rho$.  For the fully anisotropic trap, the reference frequency is taken as the geometric mean, that is, $\bar \omega = (\omega_x \omega_y \omega_z)^{1/3}$. From now, we only refer to the dimensionless variables.

We perform the numerical simulations of the 3D GP equation~(\ref{gpe3d}) using the split-step Crank-Nicolson method 
described in Refs. \cite{Gammal2006,CPC1}. The dipolar integral in equation~(\ref{gpe3d}) diverges at a short distance 
in coordinate space. However, this can be circumvented by evaluating the integral in momentum space~\cite{Goral2002,CPC2}.
 For the parameters we used in this work for the dipolar GP with ($a=0$), there is no minimizer \cite{ Bao2010, Carles2015}, which means there is no actual ground state. So the states we refer here are in fact lowest local minimum states.

 The numerical simulations for the cylindrically symmetric and fully anisotropic traps are carried out 
with $192\times192\times192$ and $320\times128\times128$ grid sizes respectively, with $\Delta x = 
\Delta y = \Delta z = 0.2$ (space step) and $\Delta t = 0.004$ (time step). 
\section {Non-vortex and Vortex states Stability} \label{sec:stab} 
\begin{figure}[htpb]%
\begin{center}
\includegraphics[width=0.49\columnwidth]{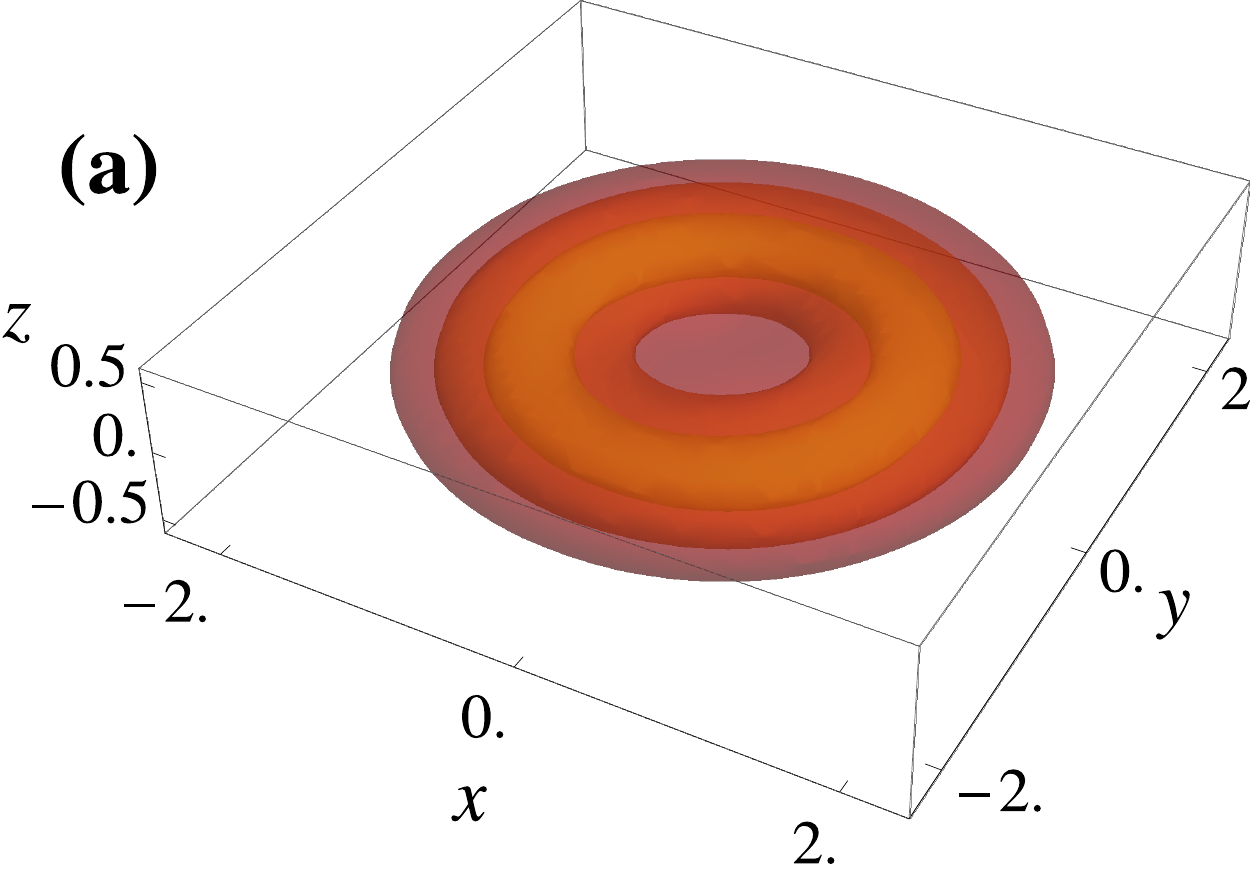}
\includegraphics[width=0.49\columnwidth]{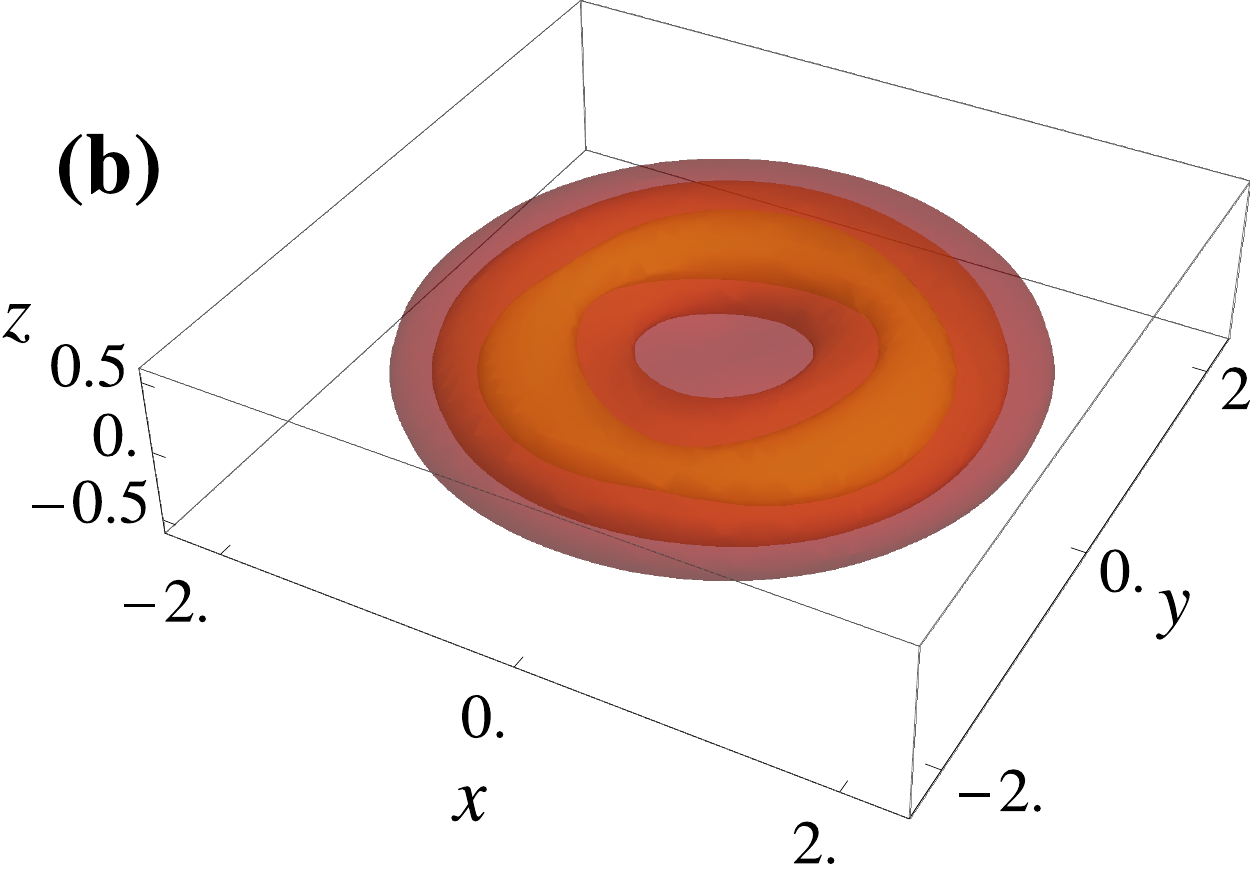}
\end{center}
\caption{(Color online) Three-dimensional contour plot showing the transparent view of density oscillation of biconcave shaped condensate for $\lambda = 7$ and $D = 30.4$ in the absence of rotation ($\Omega=0$) evolve in real-time at time, (a) $t=0$ and (b) $t=500$. The contours levels are taken as $\vert \phi({x,y,z}) \vert^2 = 0.03$ and $0.04$.}
\label{fig1}
\end{figure}

The important feature of dipolar BEC is the emergence of unusual states. 
The lowest local minimum state looks like a biconcave shaped structure where the density maximum  
is not in the center of the condensate~\cite{Ronen2007,Dutta2007}. Furthermore, the density 
oscillations of dipolar BEC with two and four peaks were also reviewed in a fully 
anisotropic trap. In this section, we are interested in  studying the stability and structures 
in purely dipolar BEC in pancake trap. We have composed the stability diagram for a purely dipolar BEC 
as a function of the trap aspect ratio ($\lambda$) and dipolar interaction strength ($D$). A similar 
stability diagram is given in Ref.~\cite{Ronen2007}. A dipolar BEC is unstable and collapses for the 
number of atoms $N$ above a critical value. This can be studied by solving the three-dimensional 
time-dependent GP equation~(\ref{gpe3d}). We have broadened the stability diagram to $\lambda=30$ and 
calculated the stability region of the solutions. The biconcave shaped condensate density as shown 
in figure~\ref{fig1}(a) is obtained for the parameters $\lambda = 7$ and $D = 30.4$ from imaginary 
time propagation. In figure~\ref{fig1}(b) we have shown the real-time evolution of the 
biconcave shaped state. One may note that similar biconcave structures are shown in Refs.~\cite{Ronen2007, Dutta2007}. 
Also, we observed such local density fluctuation for the following trap aspect ratios $\lambda = 7,8,11,12,15,16,19,$ and $20$. The structured condensates are locally stable in the sense that they are stable only within a local minimum of the energy. 
\begin{figure}[htpb]
\begin{center}
\includegraphics[width=0.35\columnwidth]{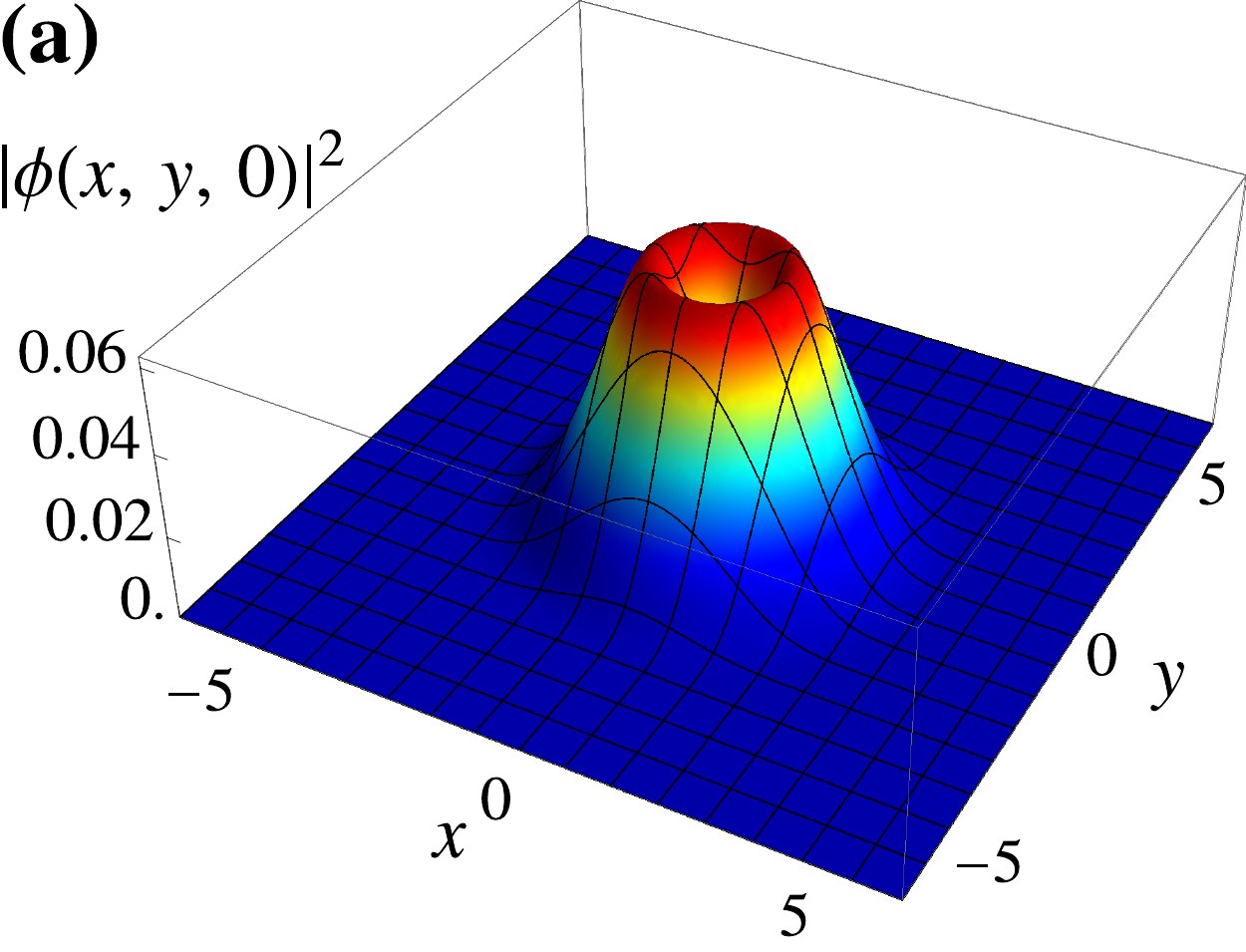}
\includegraphics[width=0.35\columnwidth]{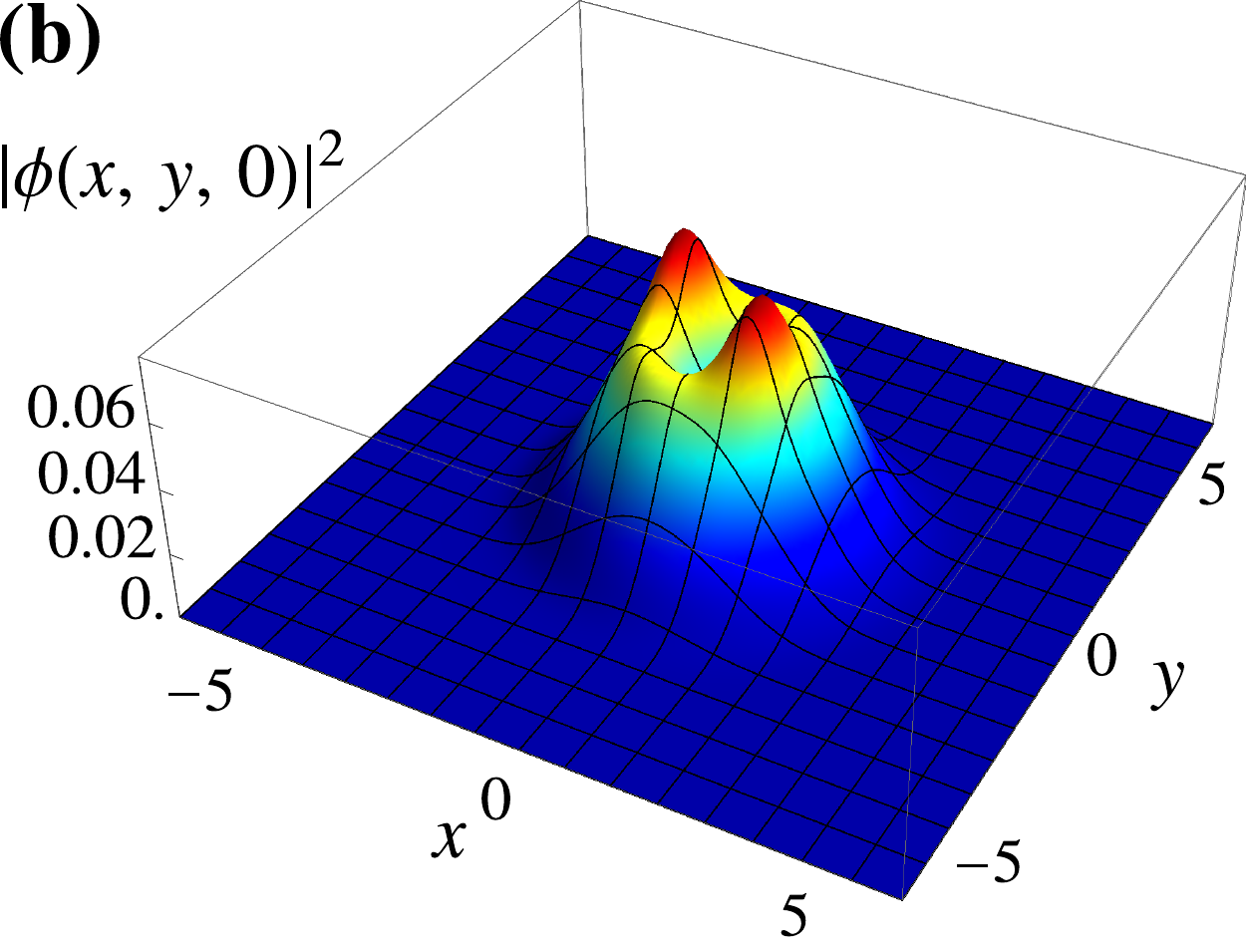}
\includegraphics[width=0.35\columnwidth]{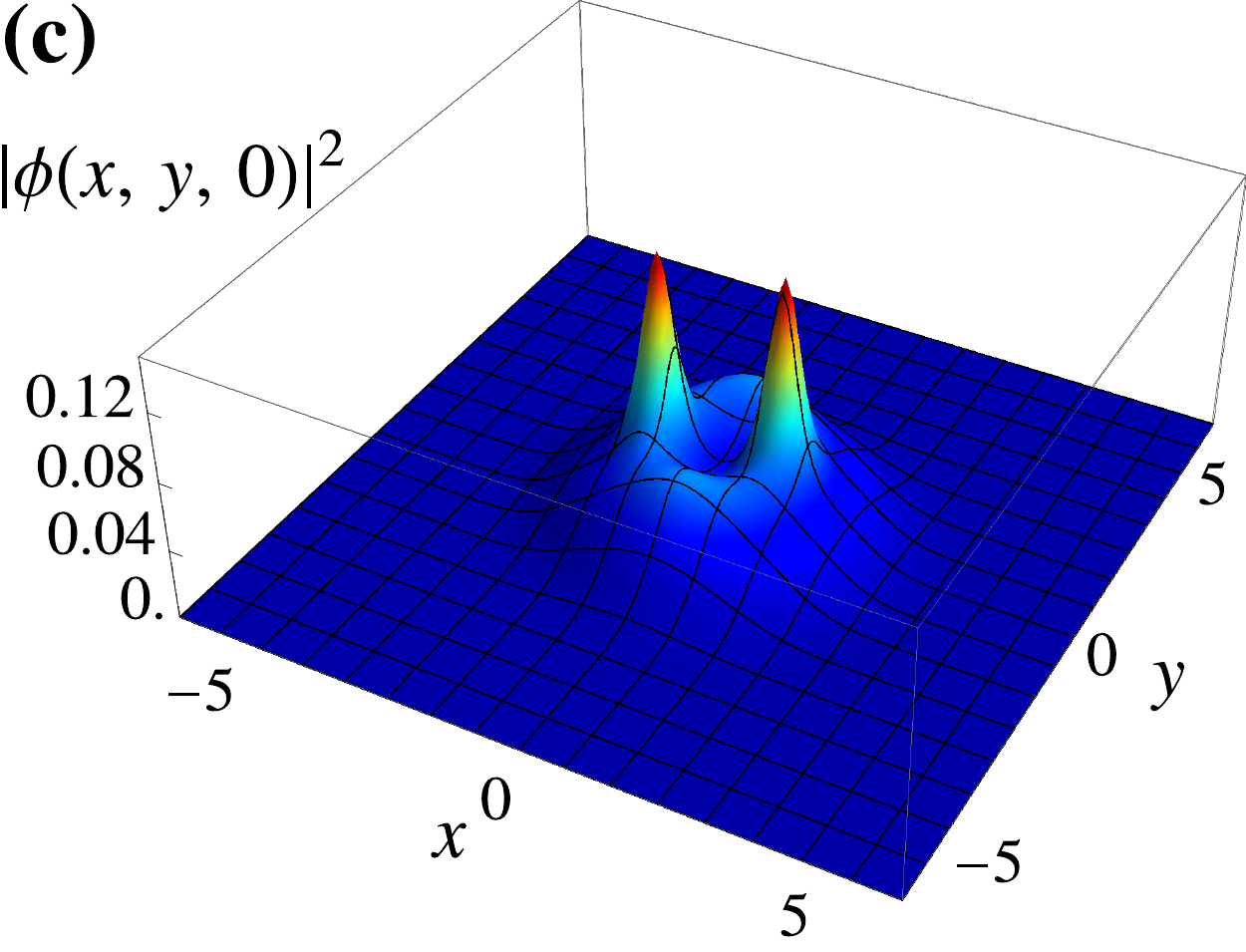}
\includegraphics[width=0.35\columnwidth]{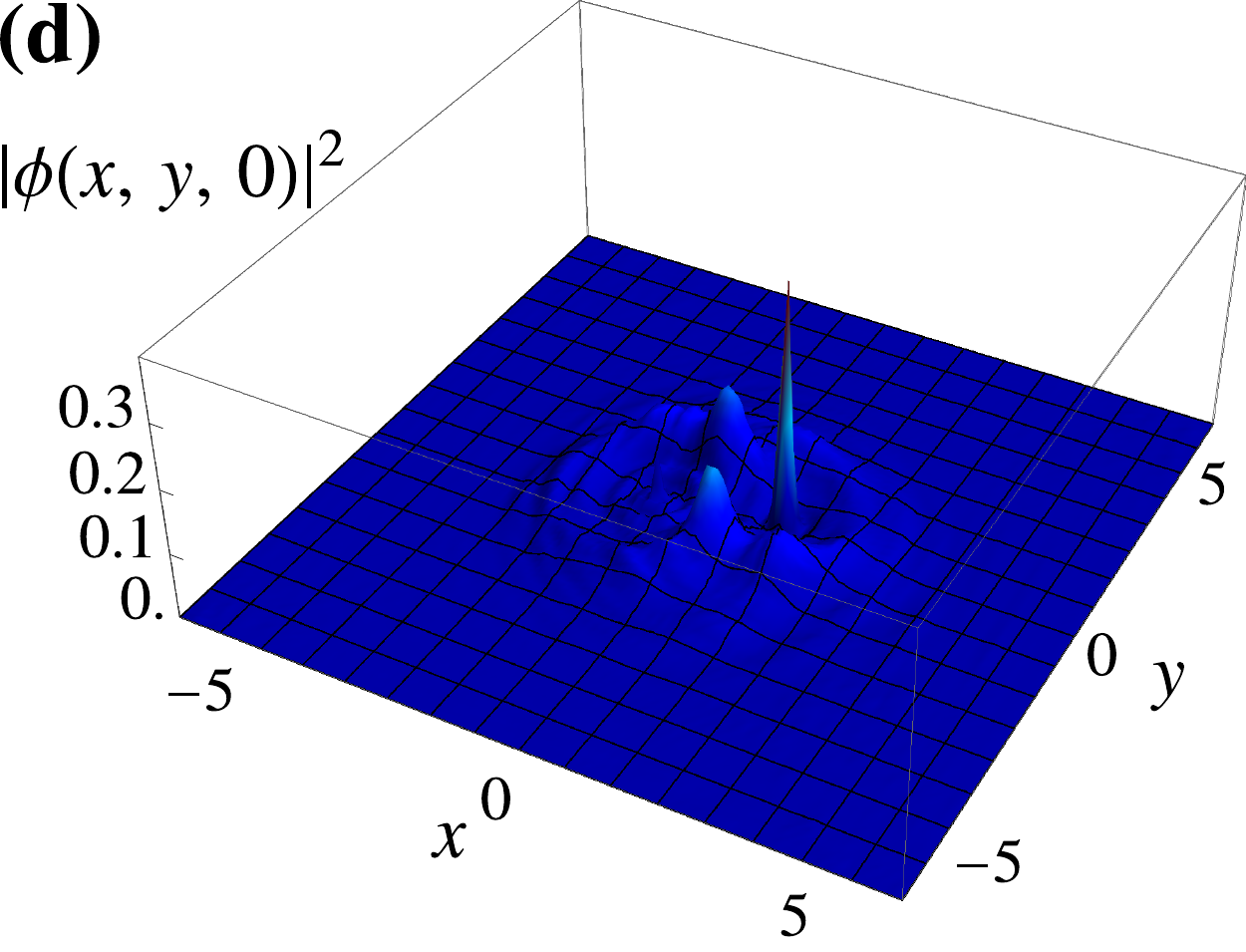}
\end{center}
\caption{(Color online) The condensate density $\vert\phi({x,y,0})\vert^2$ for $\lambda = 7$ and $D = 30.4$ in $xy$ plane showing the collapse dynamics of rotating biconcave shaped condensate with $\Omega = 0.5$ in real-time at time, (a) $t=0$, (b) $ t=60$, (c) $t=64$, and (d)  $t=65$.}
\label{fig2}
\end{figure}%
Further increase in the number of particles or increase in dipolar interaction energy will change the
state to unstable. Here we are interested in rotating this biconcave shaped condensate to study the time evolution. 
With this objective, we prepared the initial state solution by solving the equation~(\ref{gpe3d}) using imaginary time 
propagation in the absence of rotation ($\Omega=0$) and then seed the resulting solution in real-time propagation by 
applying a rotation with frequency $\Omega=0.5$. At the time of rotation, the dipoles are immensely pulled towards the 
outer rim, and two peaks formed as shown in figure~\ref{fig2}(b). These two peaks sustain for a very short 
time, and biconcave shaped condensate becomes unstable.

\begin{figure}[htpb] 
\begin{center} 
\includegraphics[width=0.32\linewidth]{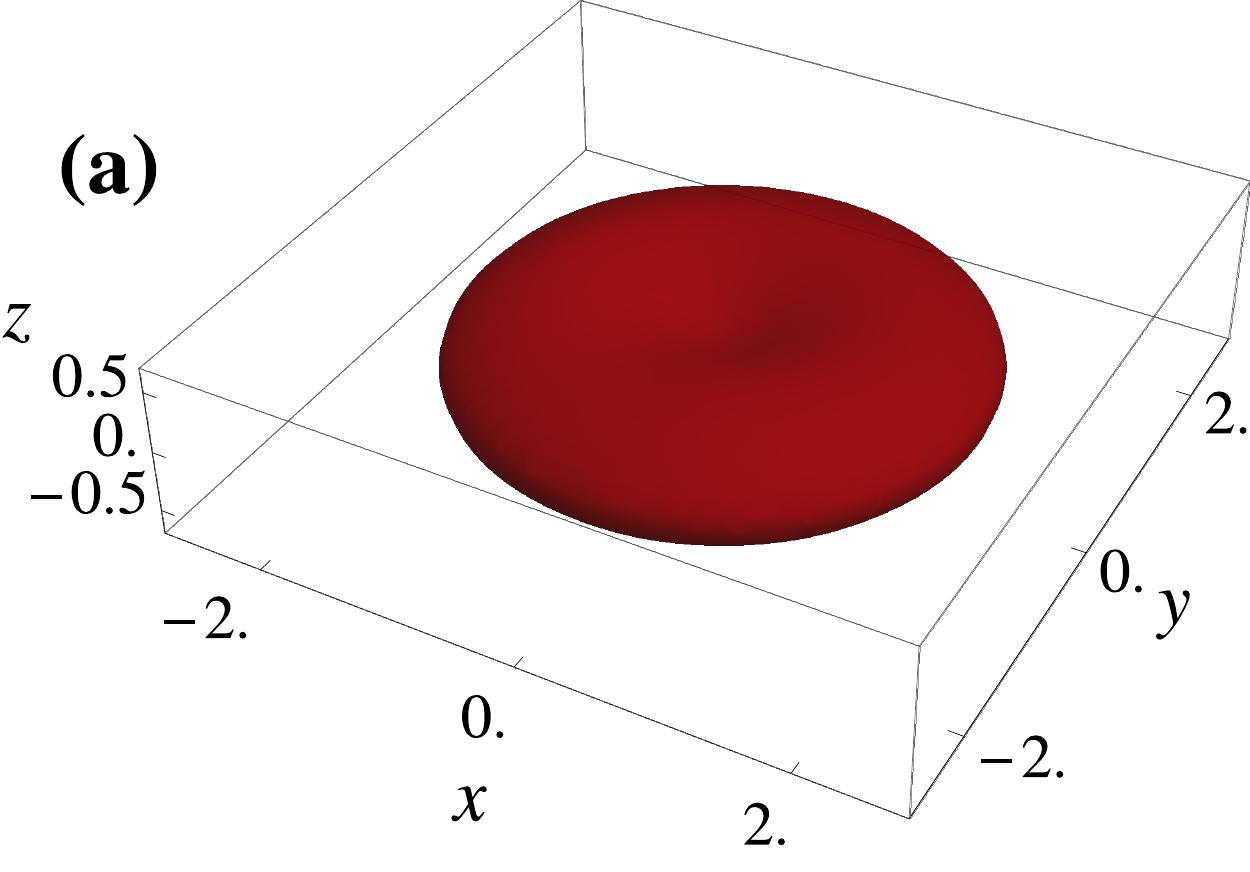} 
\includegraphics[width=0.32\linewidth]{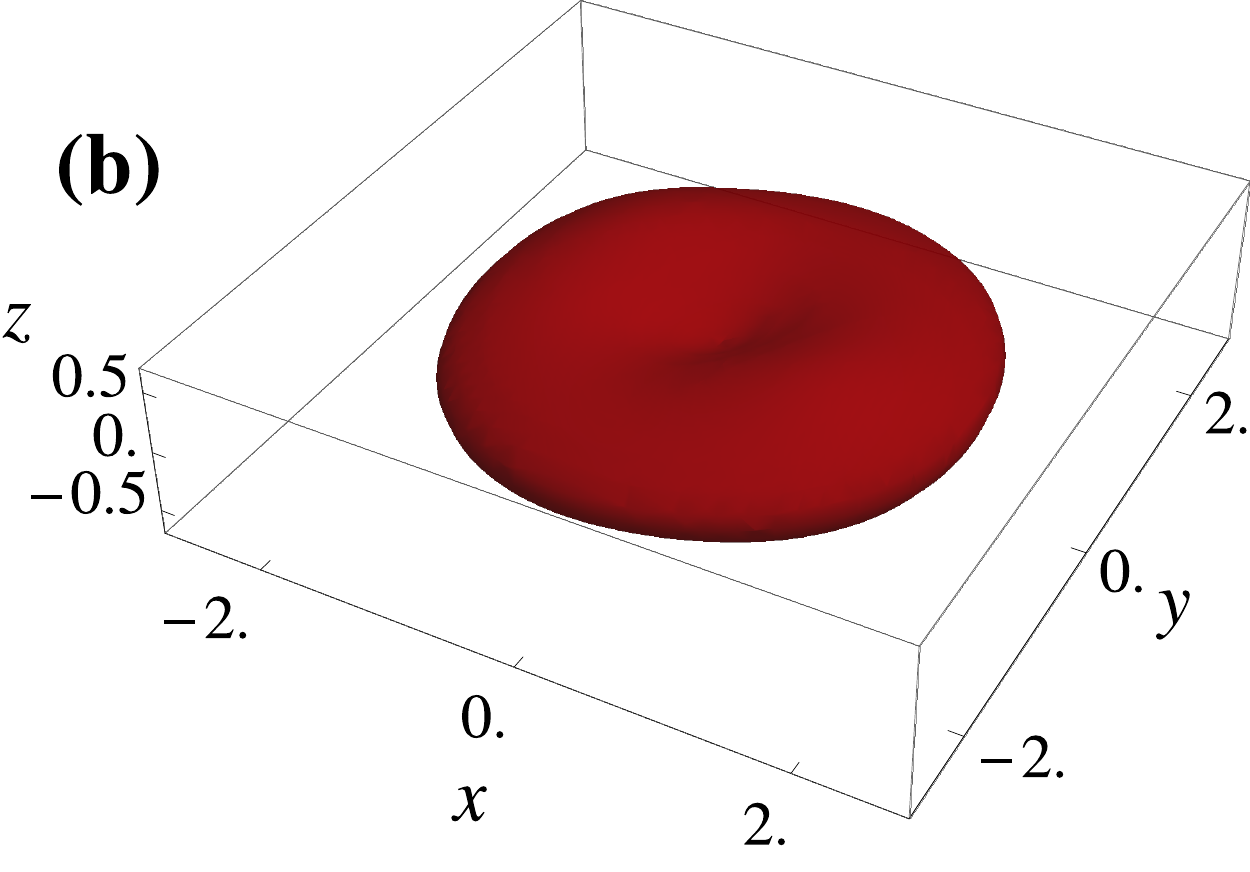} 
\includegraphics[width=0.32\linewidth]{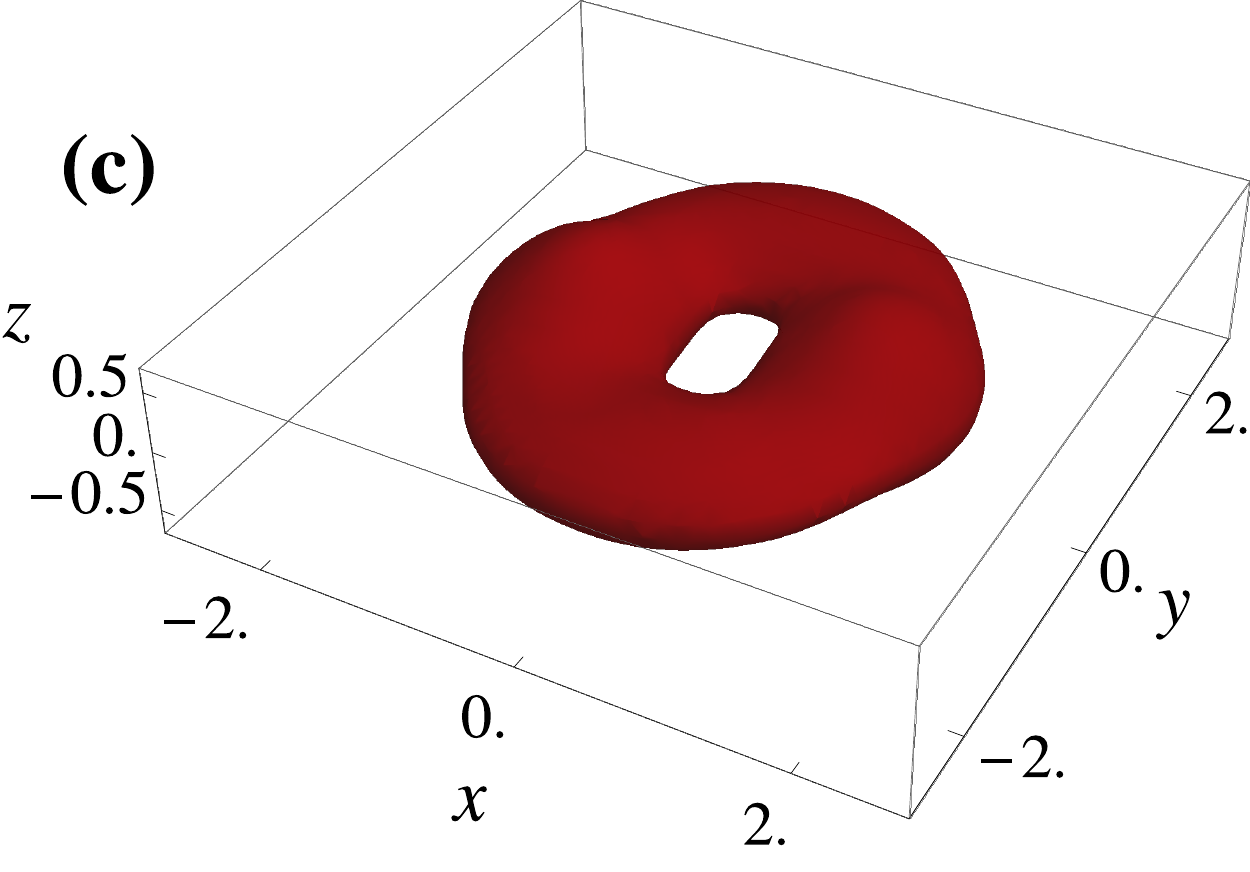} 
\end{center} 
\caption{(Color online) Three-dimensional contour plots of the 
density $\vert\phi({x,y,z})\vert ^2$ for $\lambda = 7$ and $D = 30.4$ showing the rotating biconcave shaped condensate 
with $\Omega=0.5$ in real-time propagation at times (a) $t=0$, (b) $t=60$, and (c) $t=64$. The contour levels are taken as $\vert 
\phi({x,y,z}) \vert^2 = 0.04$.} 
\label{fig3} 
\end{figure}

\begin{figure}[!ht]
\centering\includegraphics[width=0.6\columnwidth]{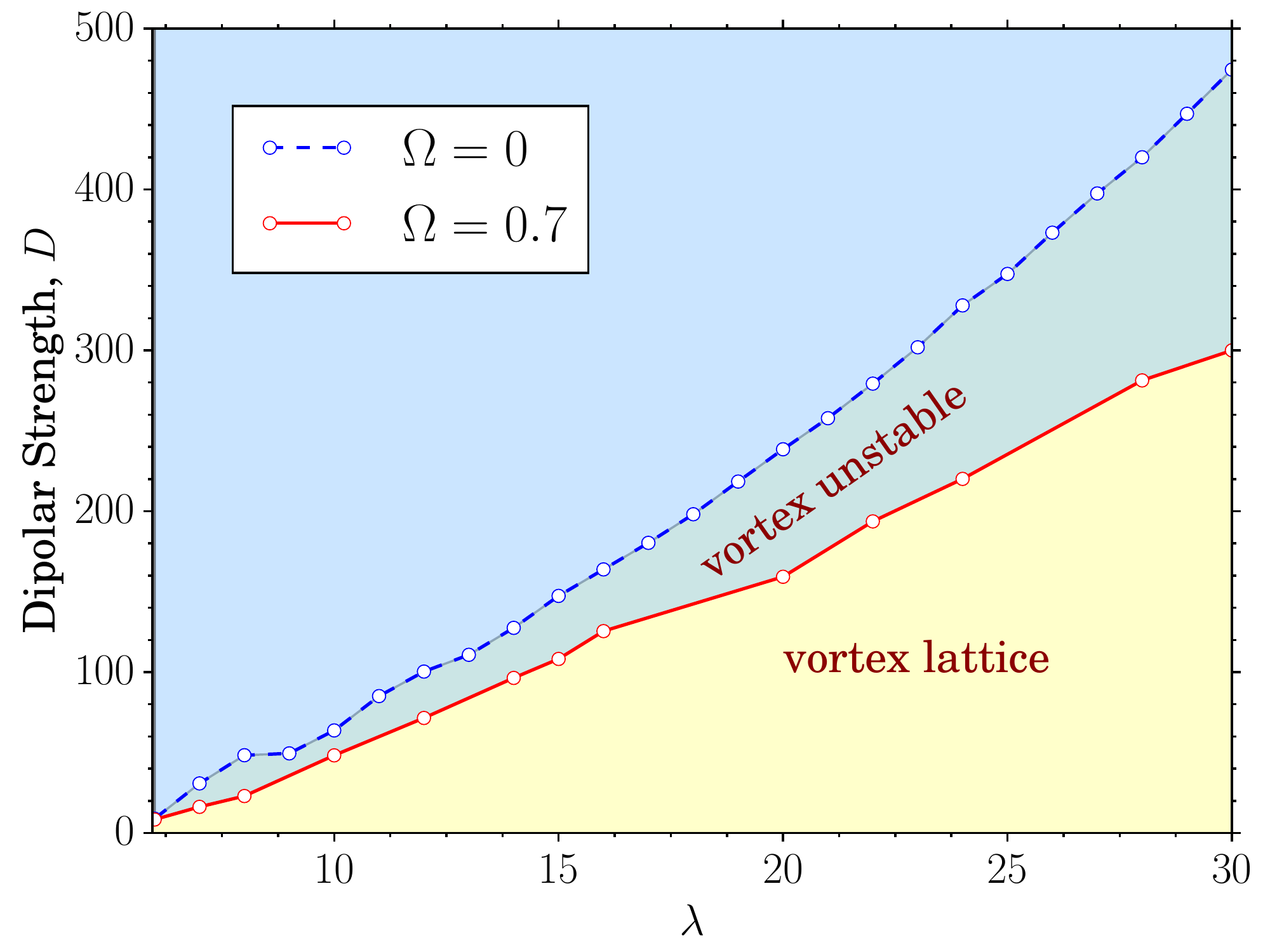}
\caption{(Color online) Stability diagram of a purely dipolar BEC in cylindrically symmetric trap as a function of trap 
aspect ratio ($\lambda$) and dipolar interaction parameter ($D$). Below the dotted blue line the local minimum 
with non-vortex state is stable while below the solid red line is the region of stable vortex lattice.}
\label{fig4}
\end{figure}

 If we analytically evolve an exact solution with zero angular momentum in the $z$ direction, then term $-\Omega L_z$ in equation 
(\ref{eqn:dgpe}) plays no role. However, this may not be the case either physically or if it is evolved numerically. Physically, 
there is always some fluctuation in the wave function field. Fluctuations are random and have in principle contributions of all 
components of different angular momentum. In the numerical calculations, roundoff errors are always present and give nonzero angular 
momentum contribution. The term $-\Omega L_z$ will then act on the perturbations, growing them exponentially 
(modulational instability) and the condensate eventually collapses. The collapse dynamics during rotation is shown as two- and three- 
dimensional view in figures~\ref{fig2} and \ref{fig3}, respectively. 
Further, we have constructed a phase diagram illustrating the 
stability region for stable vortex lattice in $\lambda-D$ parameter space as shown in figure~\ref{fig4}.
The stable region for vortex lattice is located below the stability region of the states (dashed blue line with empty 
circles) in the figure~\ref{fig4}. To carry out this observation, we have prepared the solution using imaginary 
time propagation with $\Omega=0$ and progress the solution in imaginary-time with rotation frequency $\Omega=0.7$. Nevertheless, 
the stability diagram has been checked for the rotation frequencies in the range $0.7$ and $0.99$ and no significant changes in 
the stability boundary are observed. Also, we examined the stability of stationary vortex lattice  solutions 
by evolving  in real-time. In the figure~\ref{fig4}, the region below the solid red line with empty 
circles corresponds to stable vortex lattice. The vortex states are not stable in the region between the dashed blue with empty circles and 
solid red line with empty circles. In the next section, we discuss the feasible vortex 
structures in the shaded region below the solid red line. For the present study, we mainly consider two different harmonic trap 
aspect ratios, $\lambda = 10$ and $30$, and the dipolar interaction strength is chosen within the vortex lattice 
region in the figure~\ref{fig4}.
\section{Stationary Vortex lattices in purely dipolar BECs}
\label{sec:vortex}
\begin{figure}[htpb]
\begin{center}
\includegraphics[width=0.33\columnwidth]{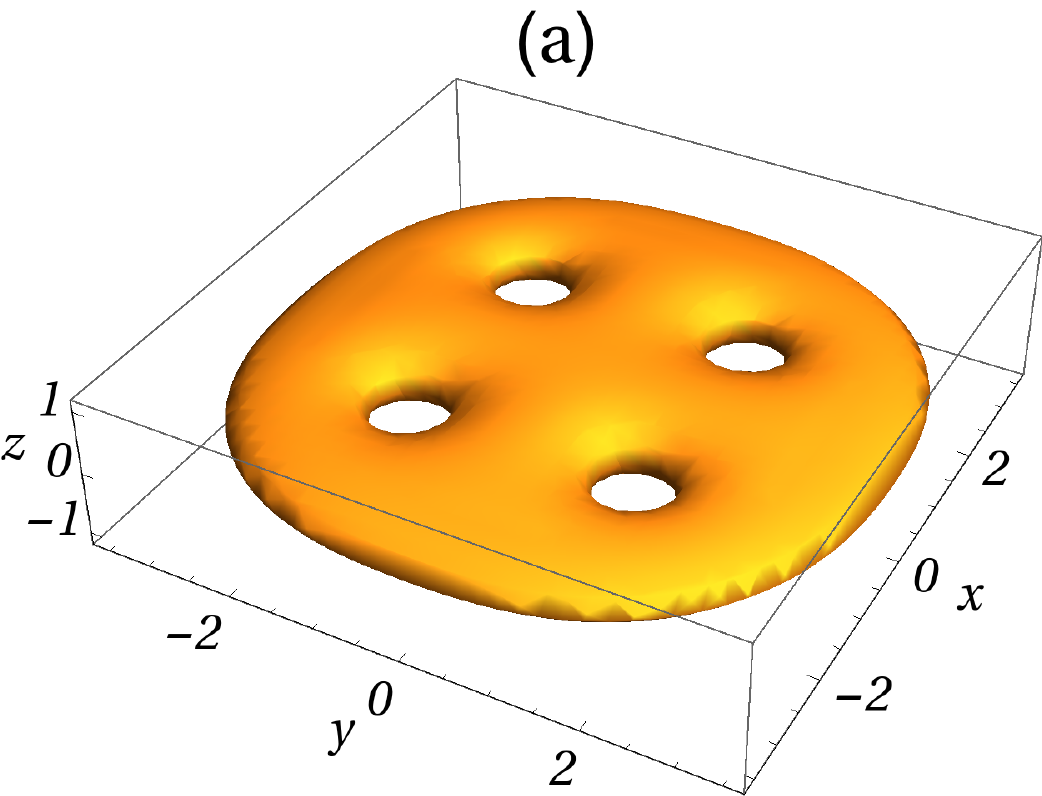}
\includegraphics[width=0.33\columnwidth]{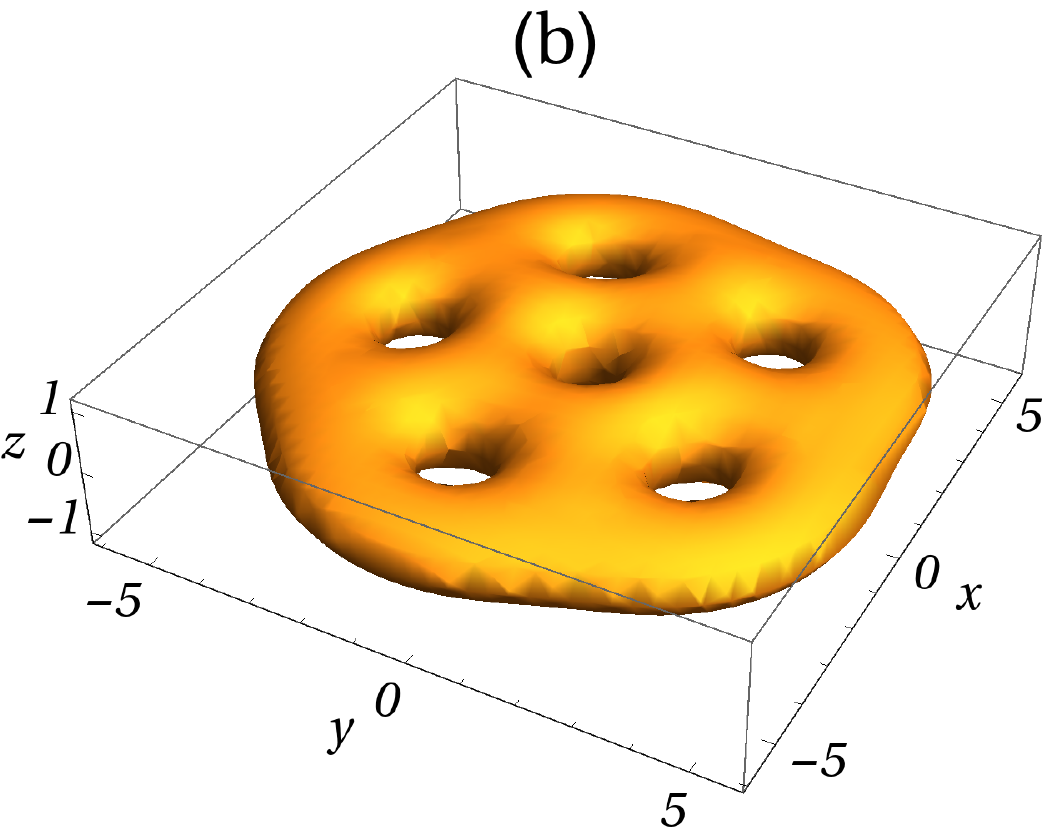}
\includegraphics[width=0.33\columnwidth]{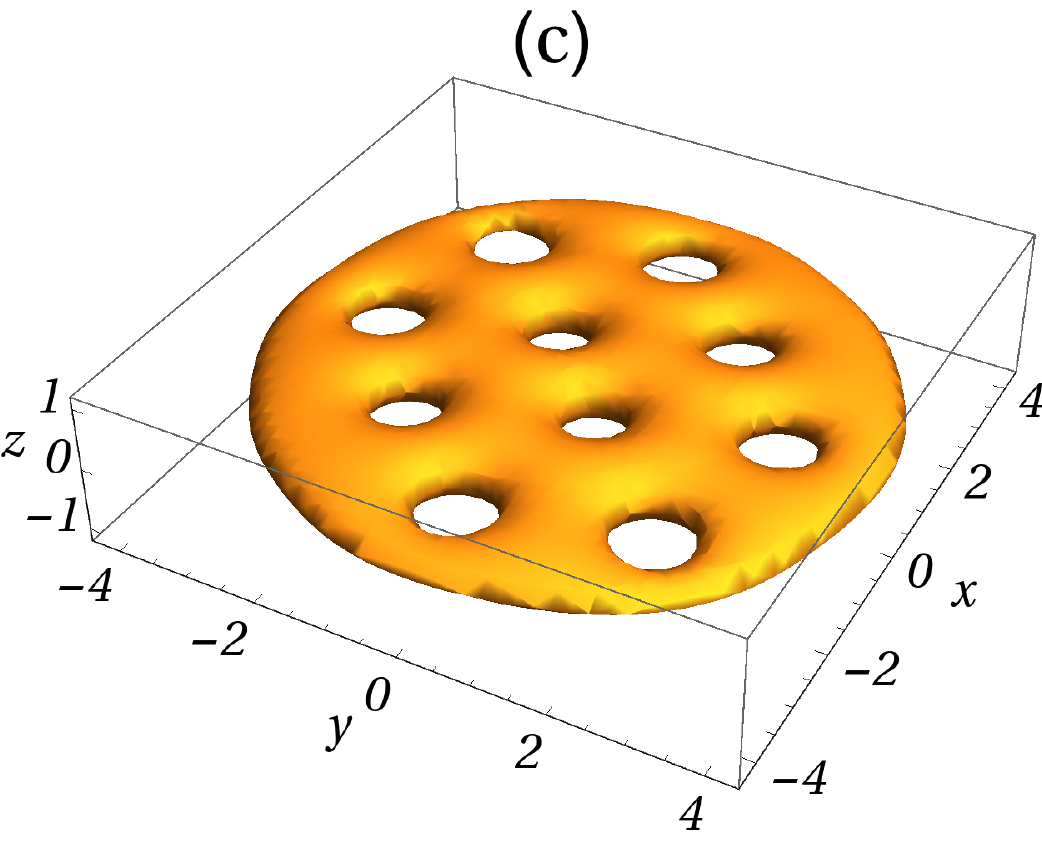}
\includegraphics[width=0.33\columnwidth]{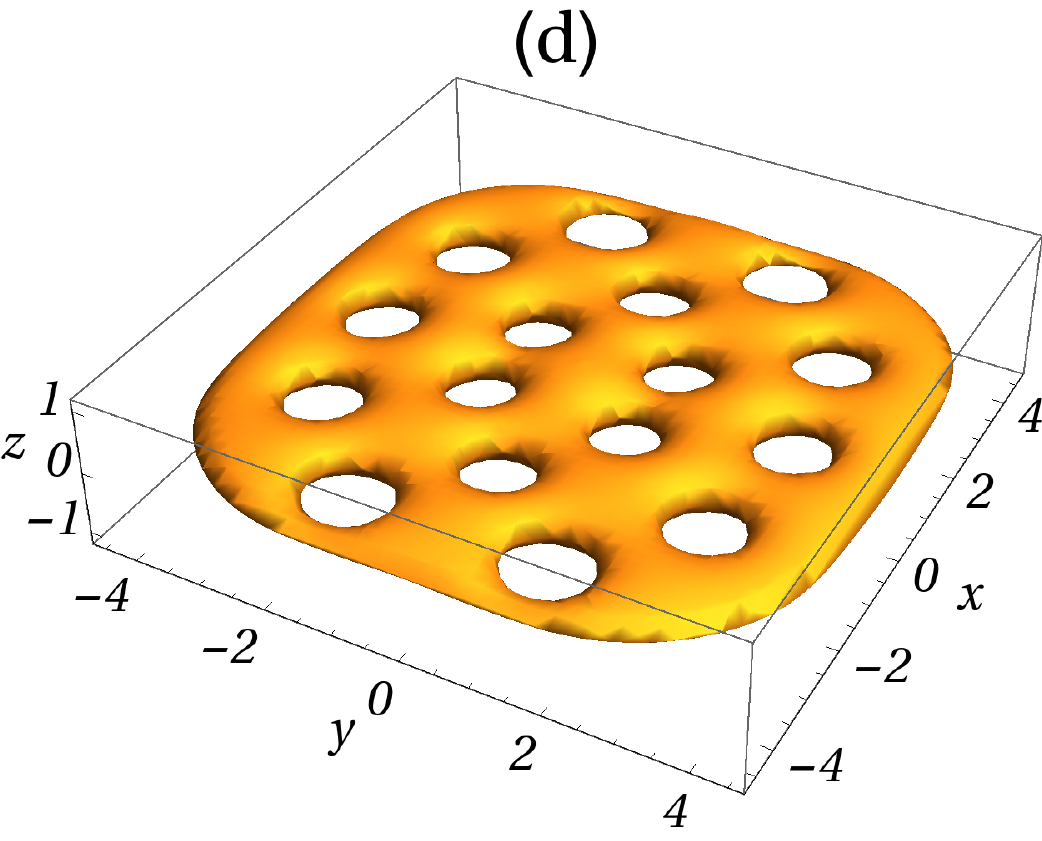}
\end{center}
\caption{(Color online) Three-dimensional view of condensate density $\vert\phi({x,y,z})\vert ^2$ with vortex lattice 
of purely dipolar BECs with $a_{dd}=16 \,a_0$, $a=0$, $D = 38$, $\lambda = 10$ at rotation frequencies 
 (a) $\Omega =0.5$, (b) $\Omega =0.7$, (c) $\Omega =0.8$ and (d) $\Omega =0.97$. The contour levels are taken as 
$\vert \phi({x,y,z}) \vert^2 = 0.0125$}
\label{fig5}
\end{figure}

In the following, we show several stationary vortex structures in purely dipolar BEC in the different harmonic trap aspect ratios 
$\lambda=10$ and $30$, and correspondingly the dipolar interaction strength is chosen as $D \simeq 38$ and $300$, respectively. 
First, in figure~\ref{fig5}, we show the stationary vortex structures in $\lambda=10$.
We prepare the initial wave function by solving equation~(\ref{gpe3d}) in the absence of rotation ($\Omega=0$). The 
vortices are then created by using imaginary time propagation with the inclusion of rotation ($\Omega \ne 0$)
 to observe the stationary vortex structures. When the condensate begins to rotate the multiply quantized vortices enter into the 
condensate from the surface. As time progress, these vortices approach to a stationary vortex configuration as shown 
in figure~\ref{fig5}.
\begin{figure}[htpb]
\begin{center}
\includegraphics[width=0.65\columnwidth]{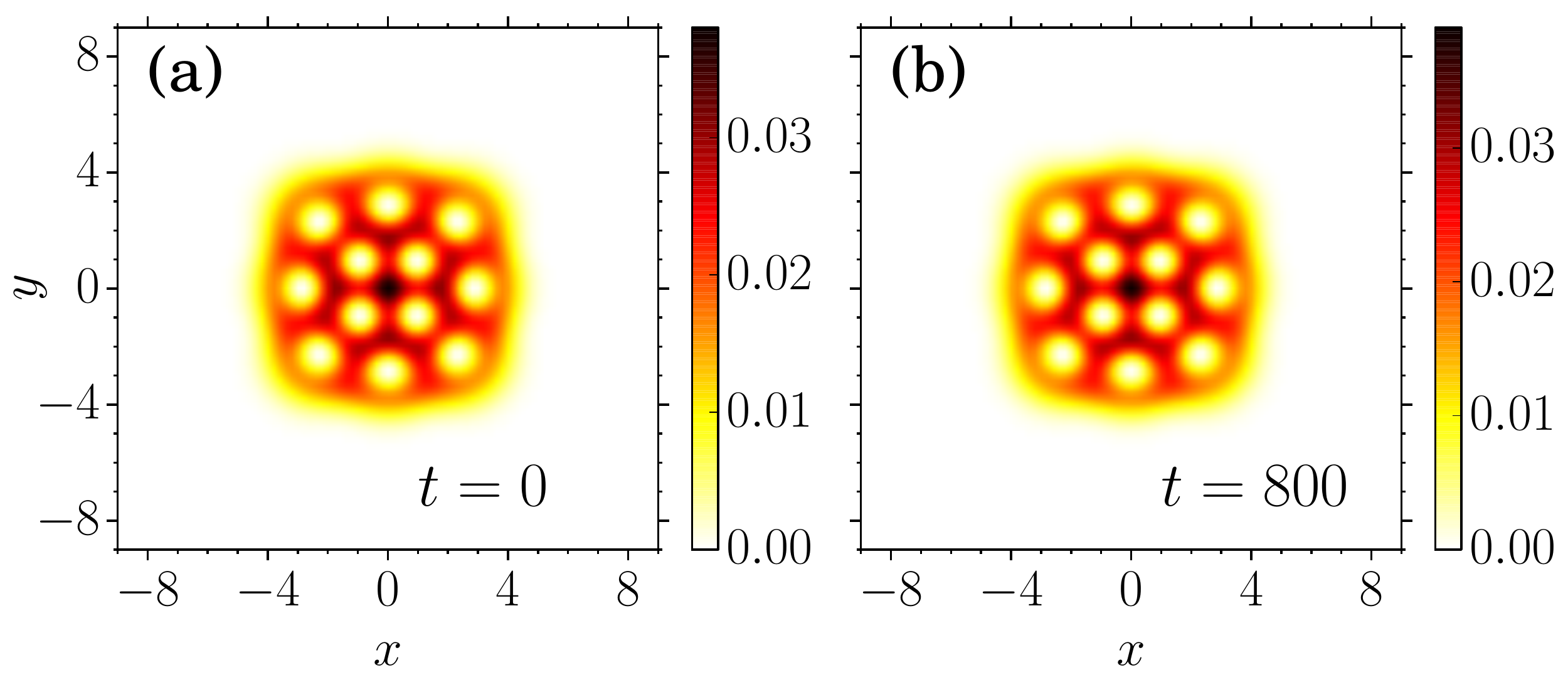}
\end{center}
\caption{(Color online) Two-dimensional view of the stable evolution of the condensate density $\vert\phi({x,y,0})\vert ^2$ with vortices arranged in a square lattice with $a_{dd}=16 \,a_0$, $a=0$, $\lambda = 10$, $D = 38$, and  $\Omega = 0.9$ at time, (a) $t = 0$ and (b) $t = 800$.} 
\label{fig6}
\end{figure}
 We notice the basic configuration of centered single vortex surrounded by five vortices at rotation frequency 
$\Omega=0.7$ as shown in figure~\ref{fig5}(b). At $\Omega=0.8$ the vortex structure is shown in figure~\ref{fig5}(c) 
resembles the structures observed in the references~\cite{Yi2006, Malet2011} for dipolar condensates. Further, we 
observed the square lattice at $\Omega=0.9$. To test the stability of the vortex structures, the {\bf relaxed} solution is 
evolved in real -time for the same parameters of imaginary time. The vortex structure persisted as depicted in 
figure~\ref{fig6}, even after long time evolution.
%
\begin{figure}[htpb]
\begin{center}
\includegraphics[width=0.33\columnwidth]{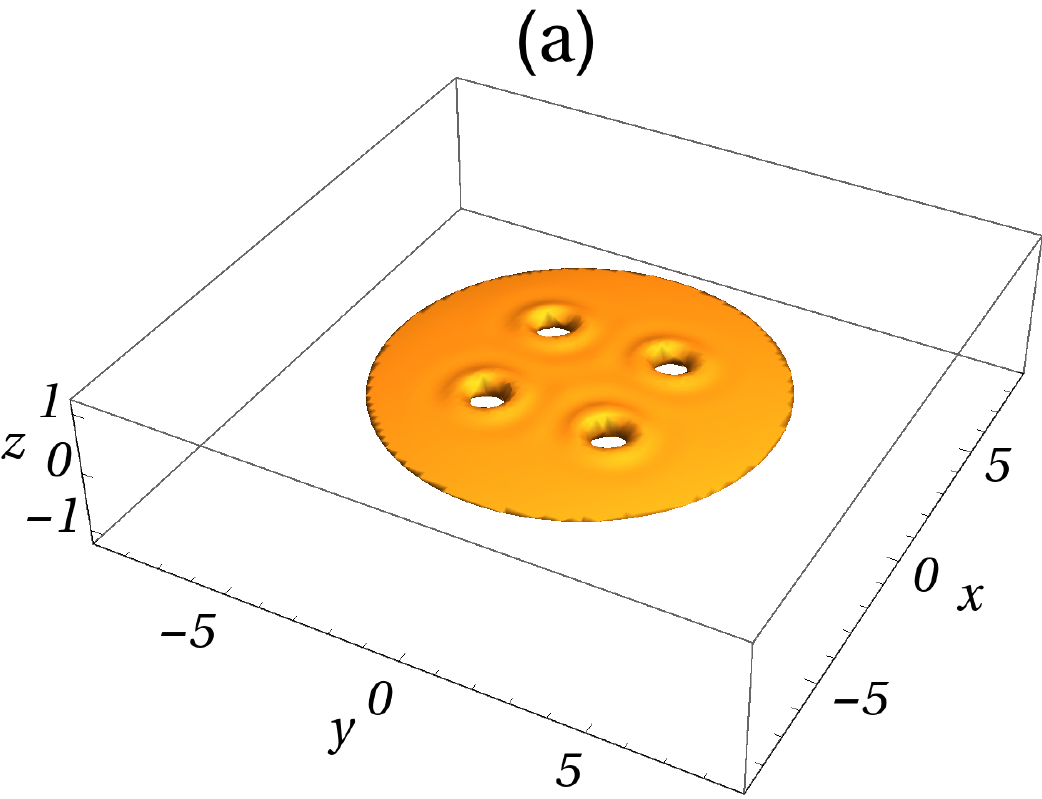}
\includegraphics[width=0.33\columnwidth]{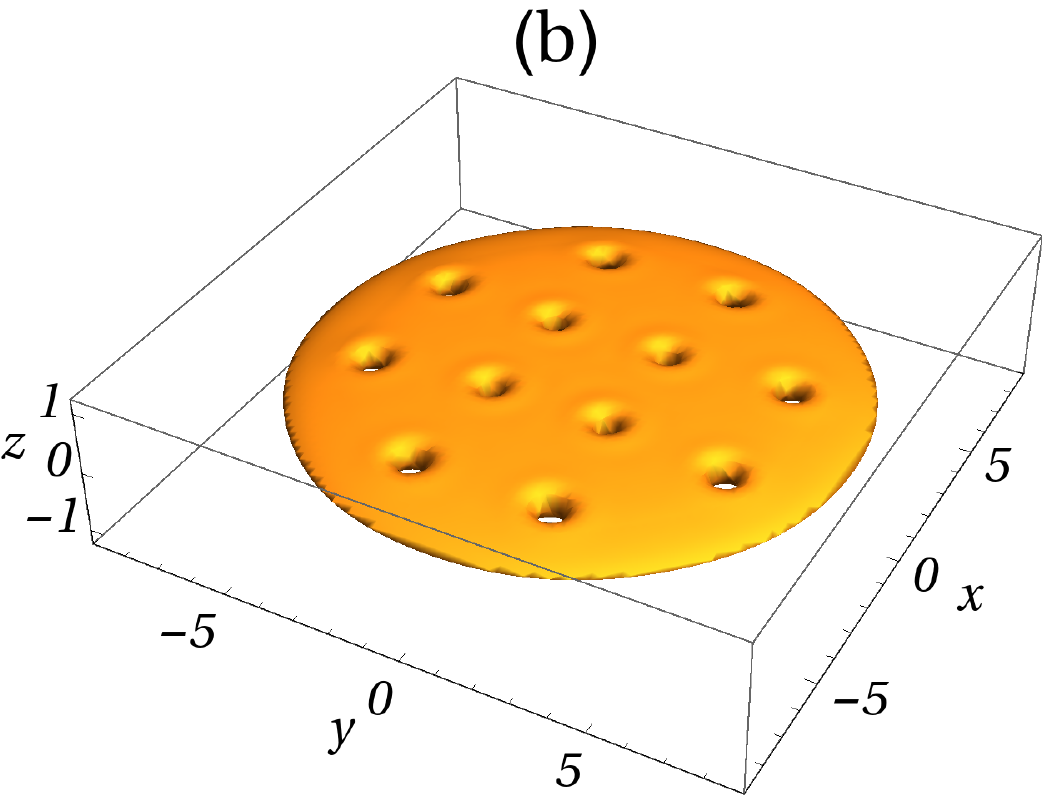}
\includegraphics[width=0.33\columnwidth]{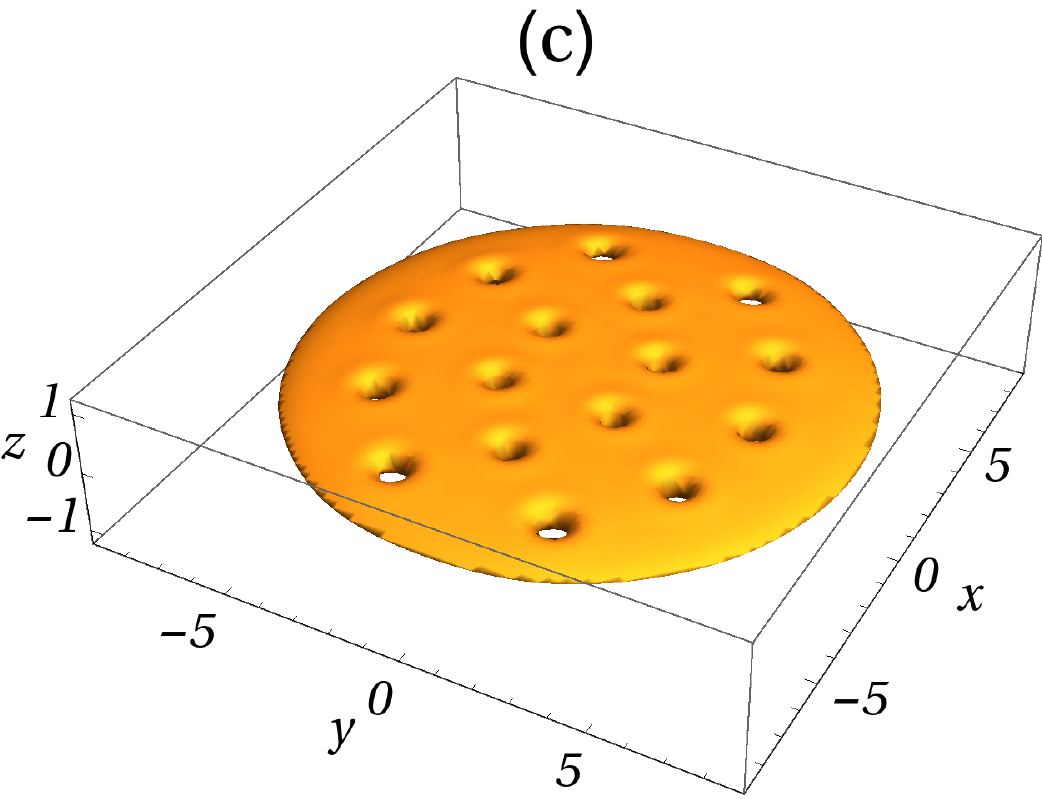}
\includegraphics[width=0.33\columnwidth]{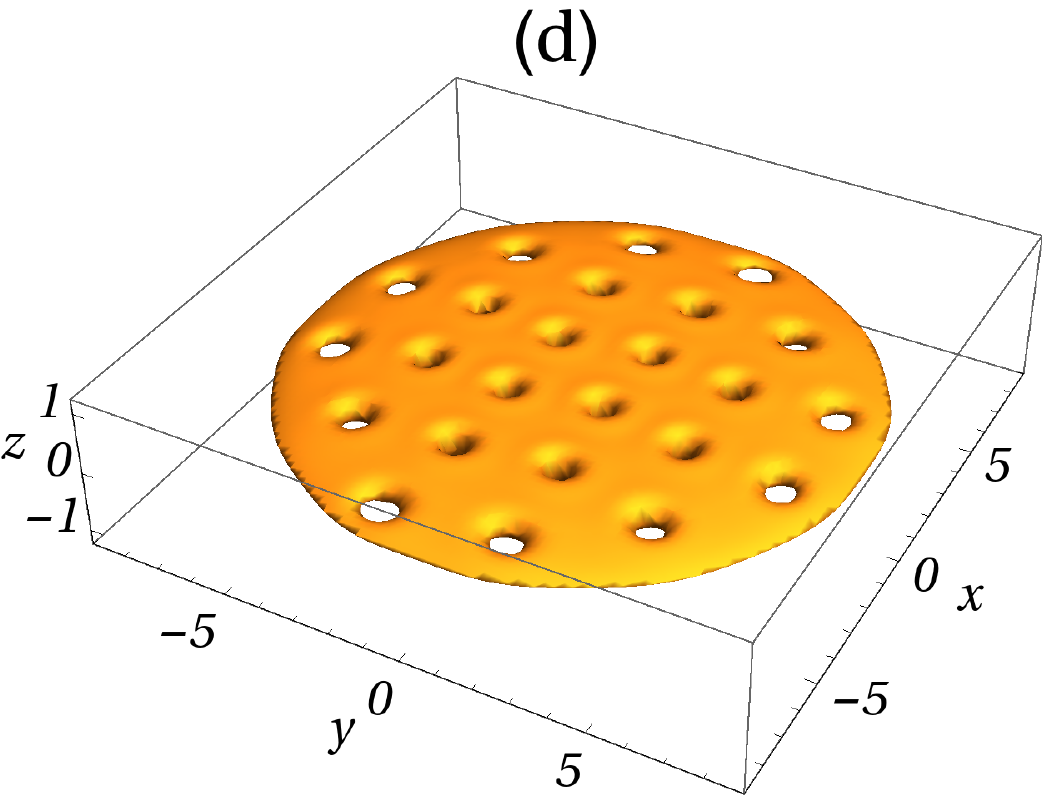}
\includegraphics[width=0.33\columnwidth]{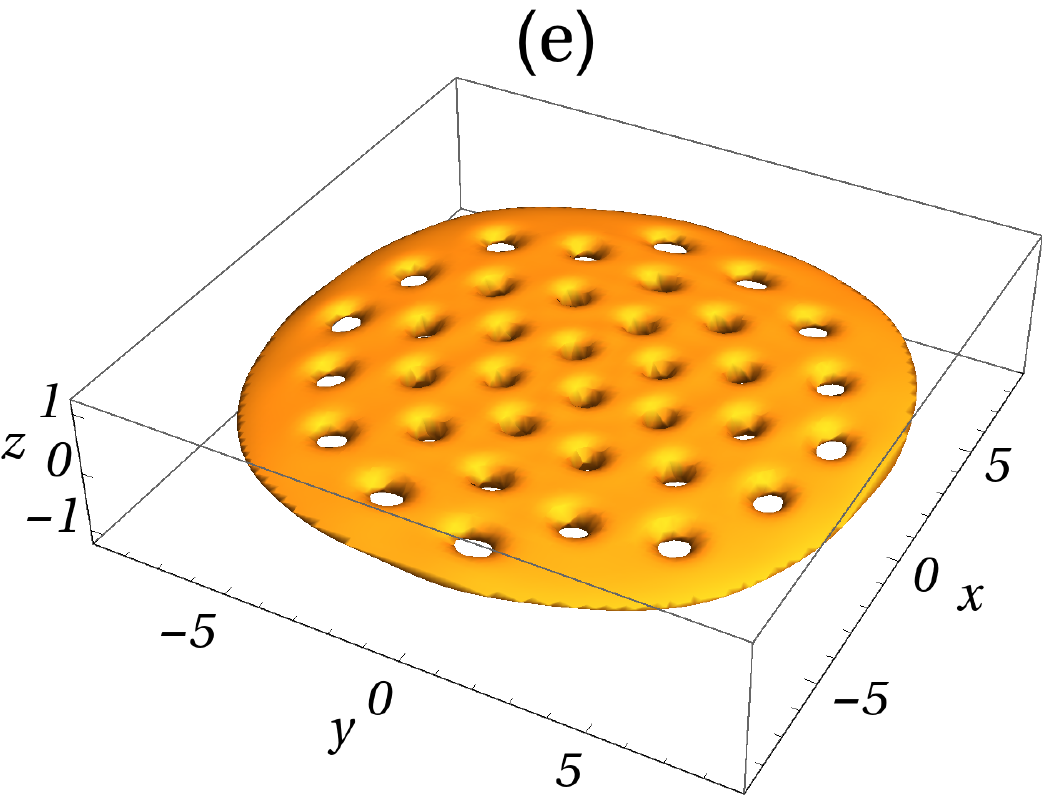}
\includegraphics[width=0.33\columnwidth]{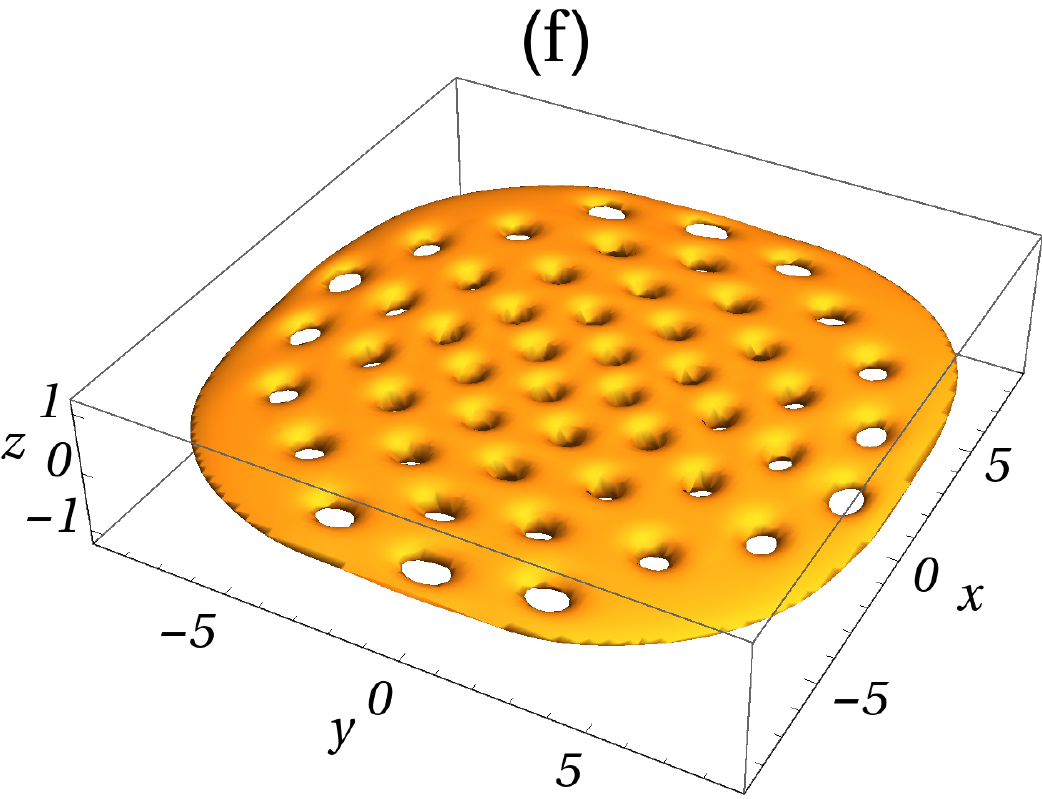}
\end{center}
\caption{(Color online)Three-dimensional view of condensate density $\vert\phi({x,y,z})\vert ^2$ with vortex lattice 
 of purely dipolar BECs with $a_{dd}=16 \,a_0$, $a=0$, $D = 300$, $\lambda = 30$ at rotation frequencies 
(a) $\Omega =0.27$, $\vert\phi({x,y,z})\vert ^2 = 0.0125$, (b) $\Omega =0.3$, $\vert\phi({x,y,z})\vert ^2 = 0.005$, (c) $\Omega =0.4$, $\vert\phi({x,y,z})\vert ^2 = 0.005$ and (d) $\Omega =0.5$, $\vert\phi({x,y,z})\vert ^2 = 0.005$ (e) $\Omega =0.8$, $\vert\phi({x,y,z})\vert ^2 = 0.005$, and (f) $\Omega =0.97$, $\vert\phi({x,y,z})\vert ^2 = 0.004$.}
\label{fig7}
\end{figure}

At high rotation frequency (e.g. $\Omega=0.97$), we found $16$ vortices as shown in figure~\ref{fig5}(d). Moreover, one may 
note from figure~\ref{fig4} that, for obtaining the vortex lattice in the condensates with larger dipolar strength, it is 
necessary to consider significantly larger $\lambda$. In this case we choose the trap aspect ratio $\lambda=30$ with dipolar 
interaction strength $D \simeq 300$ corresponding to about $118\,000$ $^{52}$Cr atoms. We observed a similar arrangement in the 
vortex structures in $\lambda= 10$ and $30$, when the number of vortices are equal. It tells us that the condensate tries to 
persist with the similar vortex structures, even though the trap aspect ratio or dipolar interaction strength are distinct. For 
instance, we noticed a qualitatively similar array of vortex lattices for both $\lambda=10$ (see figure~\ref{fig6}) and $\lambda=30$ 
(see figure~\ref{fig7}(b)), with $\Omega=0.9$ and $0.3$, respectively. We observe qualitatively similar structures for 
$\lambda=10$ (see figure~\ref{fig5}(d)) and $\lambda=30$ (see figure~\ref{fig7}(c)), with $\Omega=0.97$ and $0.4$, respectively. The 
condensate in a strong pancake trap creates a large number of vortices, even at low rotation frequency.

Further, we observed some distortion in the vortex lattice structure when the number of vortices is
sufficiently large. In particular, the distortion is larger near to the surface than near to the center of the condensate. 
All the vortices have the same charge of vorticity, and their repulsion keeps the stable configuration. 
We note that the size of the vortex core radius becomes significantly larger in the vicinity of 
the surface than in the middle of the condensate while increasing the rotation frequency. 
Consequently, near to the surface, repulsive interaction diminishes the density of vortices and leads to the 
distortion in the vortex lattice. It is equivalent to observing transverse shear waves in quantum 
gases for large rotation frequencies~\cite{Tkachen}. In this regime, it takes place a decrease in the  elastic 
shear strength of the vortex lattice.
  
\begin{figure}[htpb]
\begin{center}
\includegraphics[width=0.65\columnwidth]{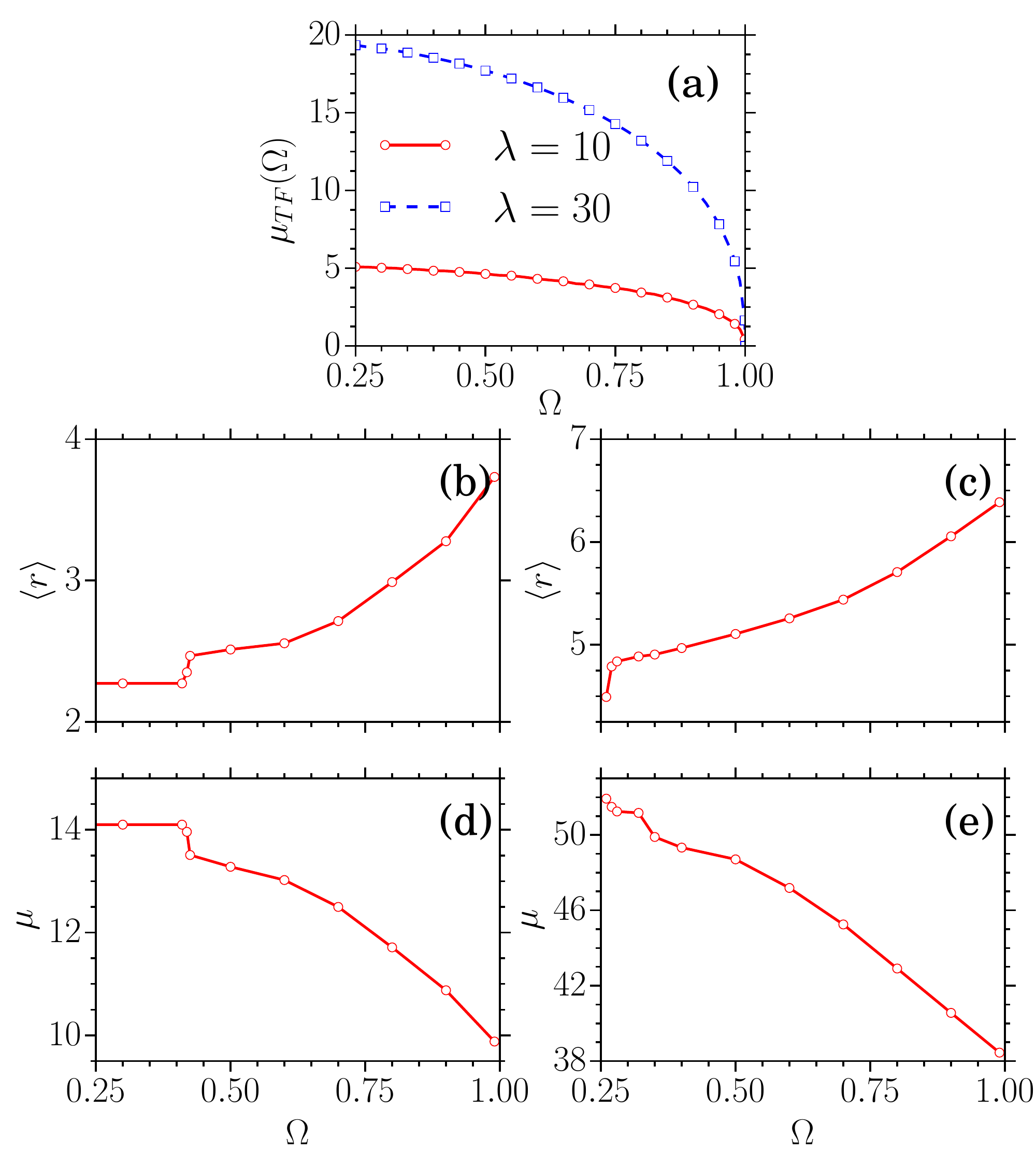}
\end{center}
\caption{Plot of the chemical potential and rms radius as a function of the rotation frequency $\Omega$ for the 
dipolar BEC with the scattering length $a=0$, $\lambda$ = 10 with $D\simeq38$ and $\lambda$ = 30 with $D\simeq300$. (a) TF chemical potential calculated from Eqn.~(\ref{chem}). Plot of the numerical calculation of radius and chemical potential in (b), (c) and (d), (e) respectively.}
\label{fig8}
\end{figure}

\begin{figure}[htpb]
\begin{center}
\includegraphics[width=0.6\columnwidth]{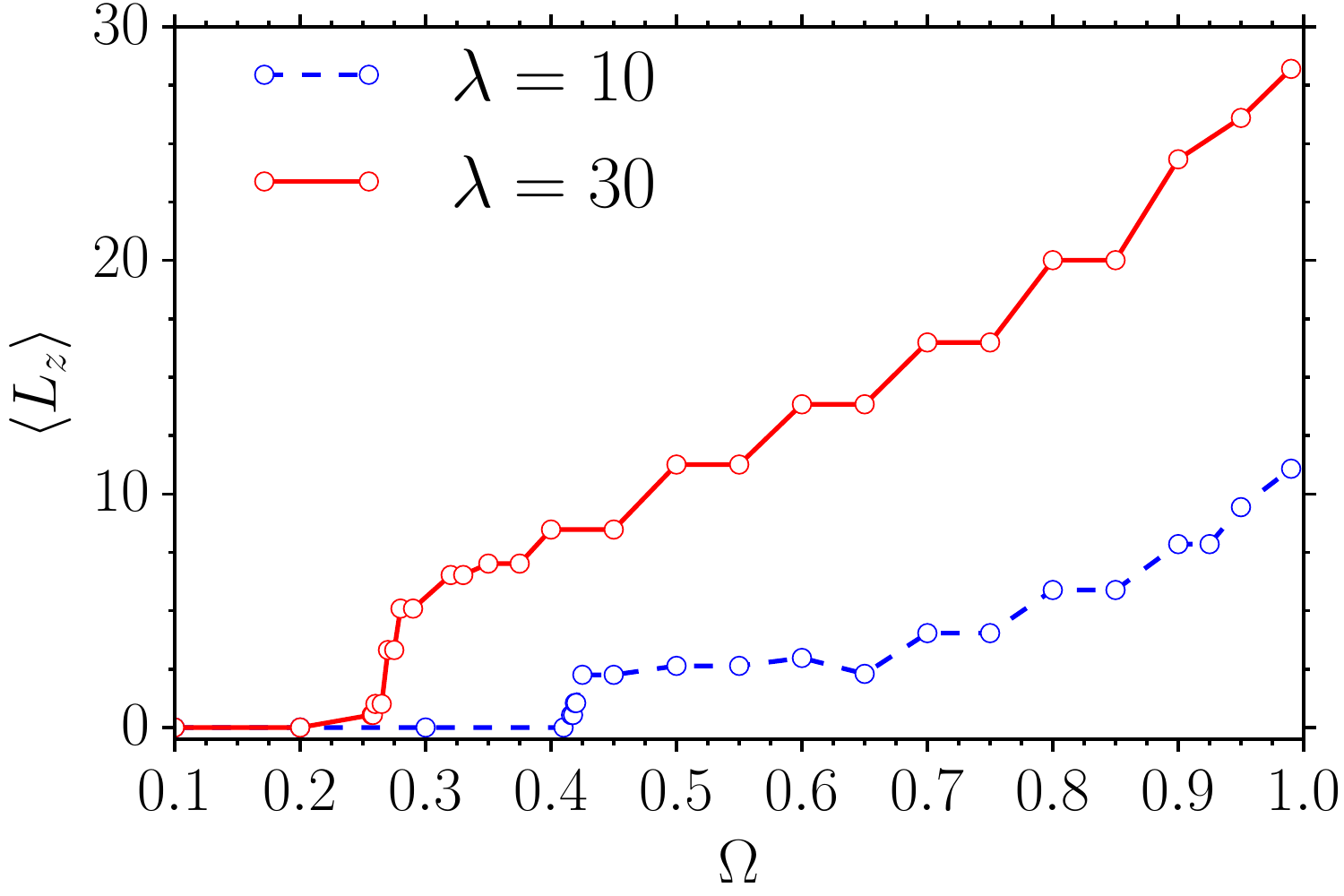}
\end{center}
\caption{(color online) Plot of the expectation value of angular momentum $\langle L_z \rangle$ as a function of the rotation frequency $\Omega$ for the purely dipolar BEC trapped in $\lambda$ = 10, 30 and $D \simeq 38$, $300$, respectively.}
\label{fig9}
\end{figure}

\begin{figure}[htpb]
\begin{center}
\includegraphics[width=0.99\columnwidth]{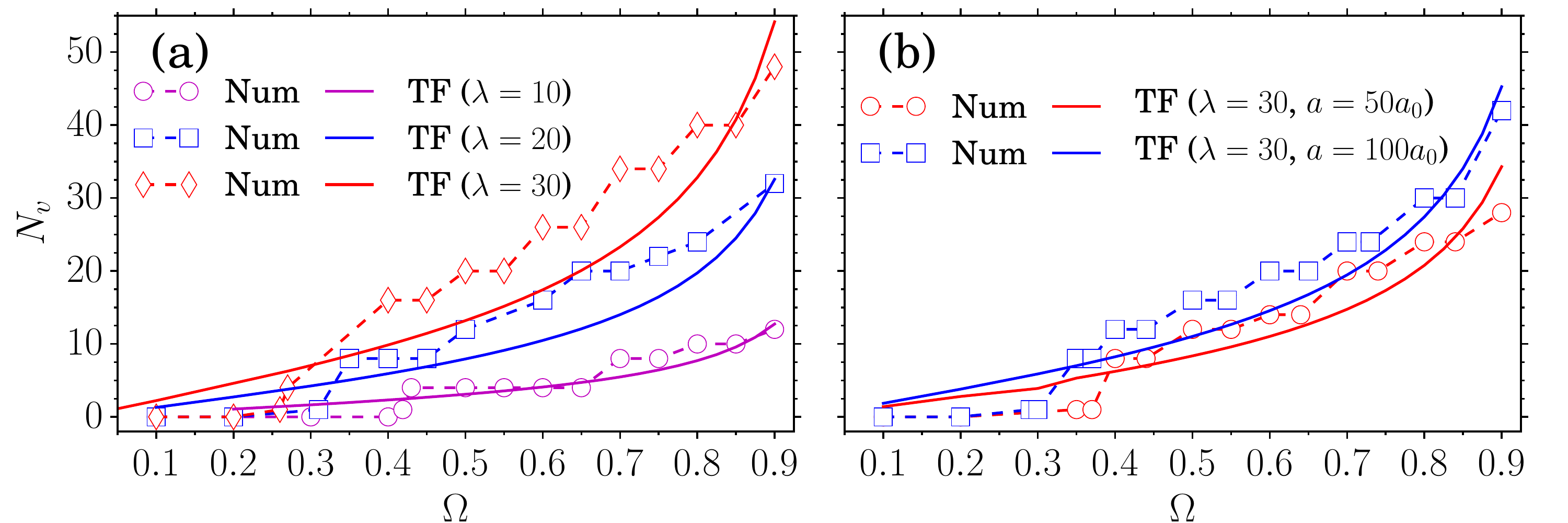}
\end{center}
\caption{(color online) Plot of the equilibrium vortex number ($N_v$) as a function of the rotation frequency $\Omega$ for the (a) purely dipolar BEC trapped in $\lambda = 10,$ $20$ and $30$, with $D=$ 38, 160 and 300, respectively and (b) non-dipolar BEC trapped in $\lambda = 30$.}
\label{fig10}
\end{figure}

In figures~\ref{fig8} and \ref{fig9}, we plot the calculated values of chemical potential, rms radius, and angular
 momentum as a function of rotation frequency for $\lambda= 10$ and $30$. First, we calculate the chemical potential with respect to the
rotation frequency in mean-field Thomas-Fermi (TF) regime. When the interaction energy is large 
compared to the kinetic energy, the kinetic energy can be neglected and enters into TF regime. We 
assume the normalized density of the dipolar BEC of the 
form~\cite{Dalfovo1999,ODell2004,Eberlein2005,Parker2008}
\begin{eqnarray}
 n({\mathbf r},t)\equiv \vert \phi({\bf r},t)\vert ^2 = \frac{15}{8\pi R_\rho^2(t) R_z(t)}\left[1-\frac{\rho^2}{R_\rho^2(t)}-
\frac{z^2}{R_z^2(t)}\right],
\label{tfden}
\end{eqnarray}
where $R_\rho(t)$ and $R_z(t)$ are the radial and axial sizes. In the TF regime one has the following set of coupled ordinary differential equations for the evolution of the condensate sizes~\cite{ODell2004}:  
\begin{eqnarray}\label{tf_dyn1} 
\ddot R_\rho & = \frac{15N}{R_\rho R_z }\left[\frac{a}{R_\rho^2}-a_{\mathrm{dd}}
\left(\frac{1}{R_\rho^2}+\frac{3}{2}\frac{f(\bar\kappa)}{R_\rho^2-R_z^2}\right)\right]  -{R_\rho}\gamma^2, \\
\ddot R_z & = \frac{15N}{R_\rho^2}\left[\frac{a}{R_z^2}+2a_{\mathrm{dd}}
\left(\frac{1}{R_z^2}+\frac{3}{2}\frac{f(\bar\kappa)}{R_\rho^2-R_z^2}\right)\right] -\lambda^{2} R_z,
\label{tf_dyn2} 
\end{eqnarray} 
with $\kappa=R_\rho/R_z$, and
\begin{eqnarray}\label{fkappa}
f(\kappa)= \frac{1+2\kappa^2-3\kappa^2d(\kappa)}{1-\kappa^2}, \;\;
d(\kappa)= \frac{\mbox{atanh}\sqrt{1-\kappa^2}}{\sqrt{1-\kappa^2}}.
\end{eqnarray}
 Here the atomic scattering length is taken as $a=0$ so that the dipolar interaction shows the dominant effect. 
It may be noted that the external rotation expands the condensate radially and shrinks it axially. As a consequence, the 
dependence of TF radii on $\Omega$ can be given by~\cite{review},
\begin{eqnarray}
\frac{R_\rho(\Omega)}{R_\rho(0)} = \left(1-\Omega^2\right)^{-3/10},\;\;
\frac{R_z(\Omega)}{R_z(0)} = \left(1-\Omega^2\right)^{1/5}.
\label{radii}
\end{eqnarray}
Also, the chemical potential has the form,
\begin{eqnarray}
\mu_{TF}(\Omega) = \mu_{TF}(0)\left(1-\Omega^2\right)^{2/5}.
\label{chem}
\end{eqnarray}
 As shown in figure~\ref{fig8}(a) the chemical potential decreases continuously and it goes to zero when $\Omega=1.0$. 
In figure~\ref{fig8}(b-e) we plot the rms radius, {\bf $\sqrt{\langle r^2 \rangle} = \left(\int \phi^{\star}( {\mathbf {r}}, t)~ {\mathbf {r}}^2~ \phi( {\mathbf {r}}, t)\, d{\mathbf {r}} \right )^{1/2}$, } and chemical potential as a function of rotation frequency calculated from the numerical solution of the GP equation. As expected, the condensate rms radius increases with the increase of rotation frequency due to the expansion of condensate. The rms radius is shown in figures~\ref{fig8}(b) and (c) for $\lambda = 10$ and $30$, respectively. One may note that, in the absence of rotation, the numerically calculated chemical potential compares well with TF results~\cite{CPC2}. Whereas, in the presence of rotation, the TF chemical potential is about two times less than the numerically calculated value. In figures~\ref{fig8}(d) and (e) we show the variation of chemical potential for $\lambda = 10$ and $30$ respectively.
Next, we calculate expectation value of angular momentum  as a function of $\Omega$ from the numerical solution. Figure~\ref{fig9}, depicts the plot of $\langle L_z \rangle$ with respect to $\Omega$ for  $\lambda = 10$ and $30$. The increase in the $\langle L_z \rangle$ is associated with the entry of vortices into the condensate. We are interested in studying the dependence of the vortex number ($N_v$) on the rotational frequency $\Omega$ in purely dipolar BEC. The rotating condensate has a dense array of vortices with a uniform density  $n_v=m(\Omega \bar\omega)/\pi\hbar$. 
The number of vortices present in the condensate is~\cite{review},  
\begin{eqnarray}
N_v = \frac{m(\Omega \bar\omega)}{\hbar} R_\rho^2(\Omega).
\end{eqnarray}
The number of vortices increases linearly with $\Omega$ assuming the BEC in a fixed axially symmetric harmonic trap. The radius $R_\rho(0)$ in TF regime can be calculated from the coupled equations~(\ref{tf_dyn1}) and (\ref{tf_dyn2}), and the corresponding $R_\rho(\Omega)$ can be obtained from equation~(\ref{radii}). In figure~\ref{fig10}(a), we have estimated the number of vortices in the different harmonic trap aspect ratios $\lambda  = 10$, $ 20$ and $30$ in TF regime and compared with numerically calculated equilibrium numbers. We noted some deviations in the number of vortices from TF results and calculated  numerically. To compare the deviation in the number of vortices in BEC without DDI, we calculate $N_v$ for the BEC without DDI by tuning to the magic angle $\varphi=54.7^{\circ}$,  where the dipolar interaction averages to zero in equation~(\ref{vdd}). Then we obtain $N_v$ at $a=50 a_0$, $100 a_0$ and we observe the deviation as shown in figure~\ref{fig10}(b).  %
 In particular, when $\Omega>0.9$ the deviation is larger and it is due to the strong expansion of 
condensate during the rapid rotation. 
\section{Vortex lattice in fully anisotropic trap} 
\label{sec:ani} 
Finally, we concerned the purely 
dipolar BEC  in fully anisotropic trap with $\gamma \neq \nu \neq \lambda$ in the 
equation~(\ref{gpe3d}). Whenever $\gamma \neq \nu$, the system breaks the cylindrical symmetry. For the 
present analysis, we fix $\lambda=10$, $\nu=1$, and vary $\gamma$ to study the changes in 
the spatial distribution of vortices in purely dipolar BECs by increasing the eccentricity of the trapping 
potential along $x$-direction. 
\begin{figure}[htpb] \begin{center} 
\includegraphics[width=0.25\columnwidth]{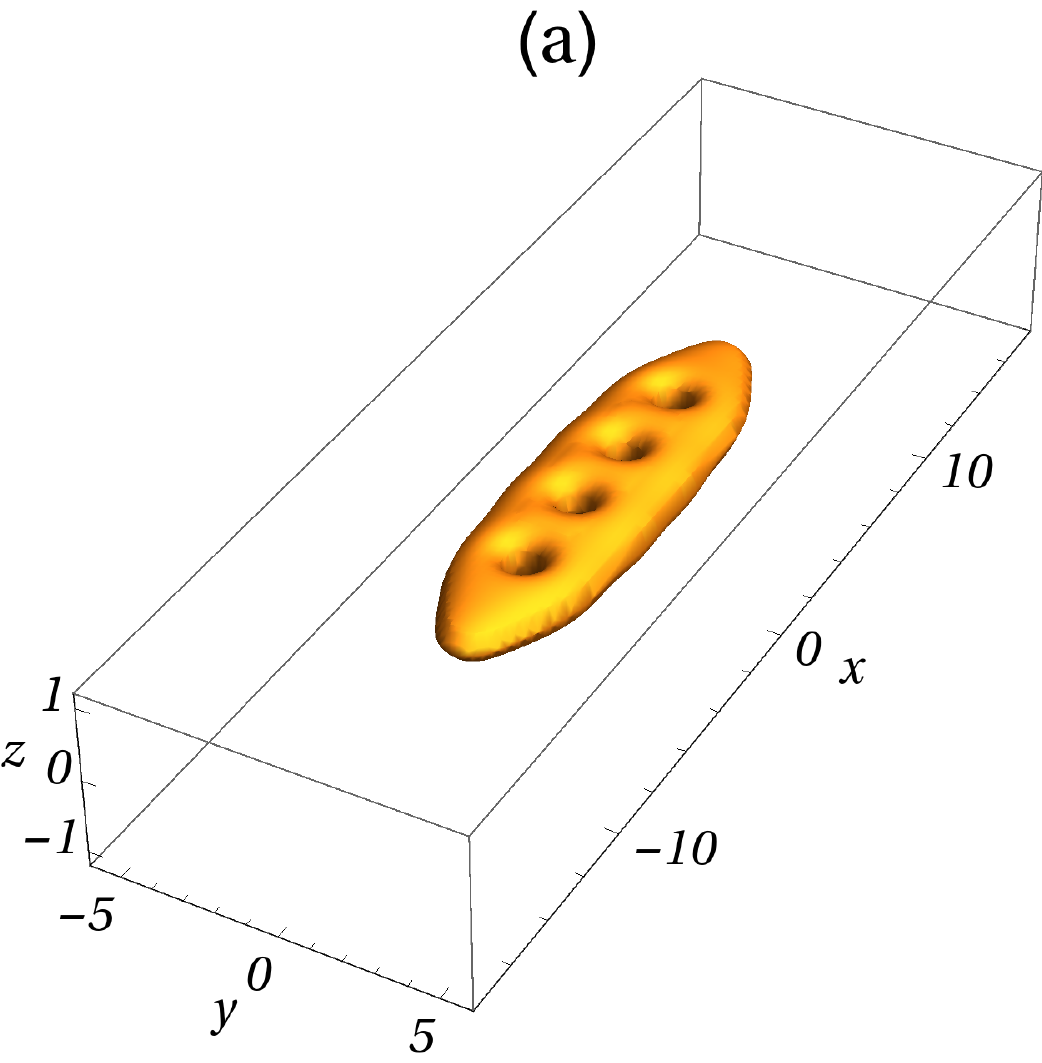}
\includegraphics[width=0.25\columnwidth]{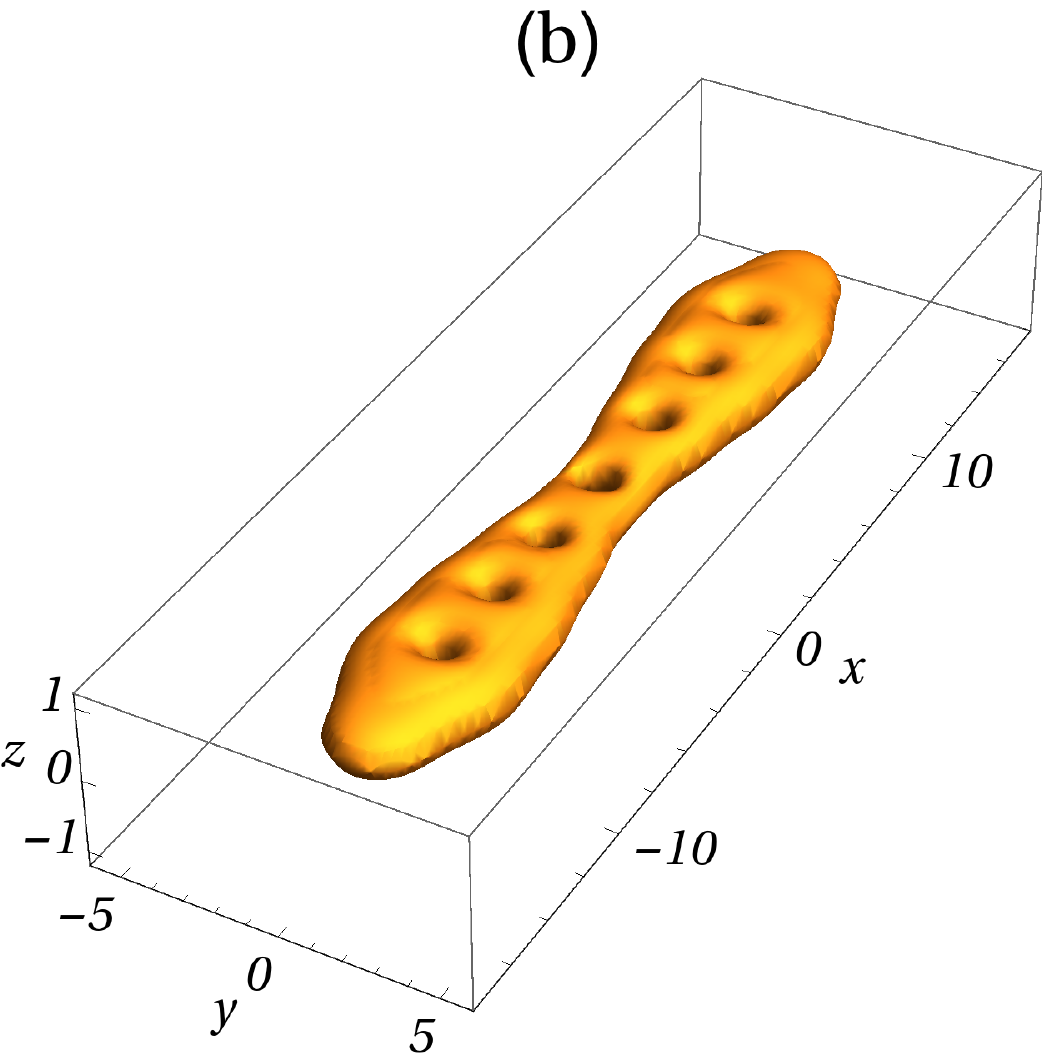} 
\includegraphics[width=0.25\columnwidth]{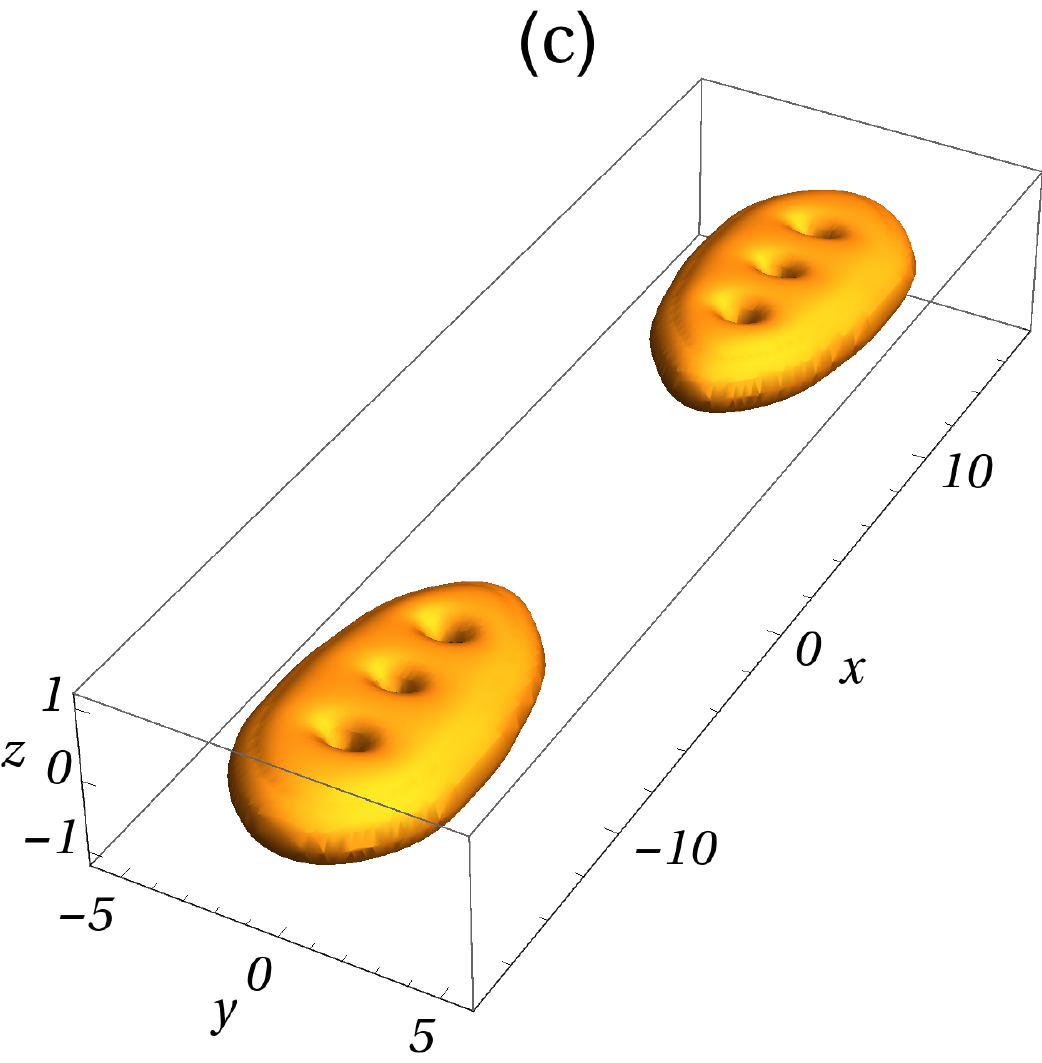} 
\includegraphics[width=0.65\columnwidth]{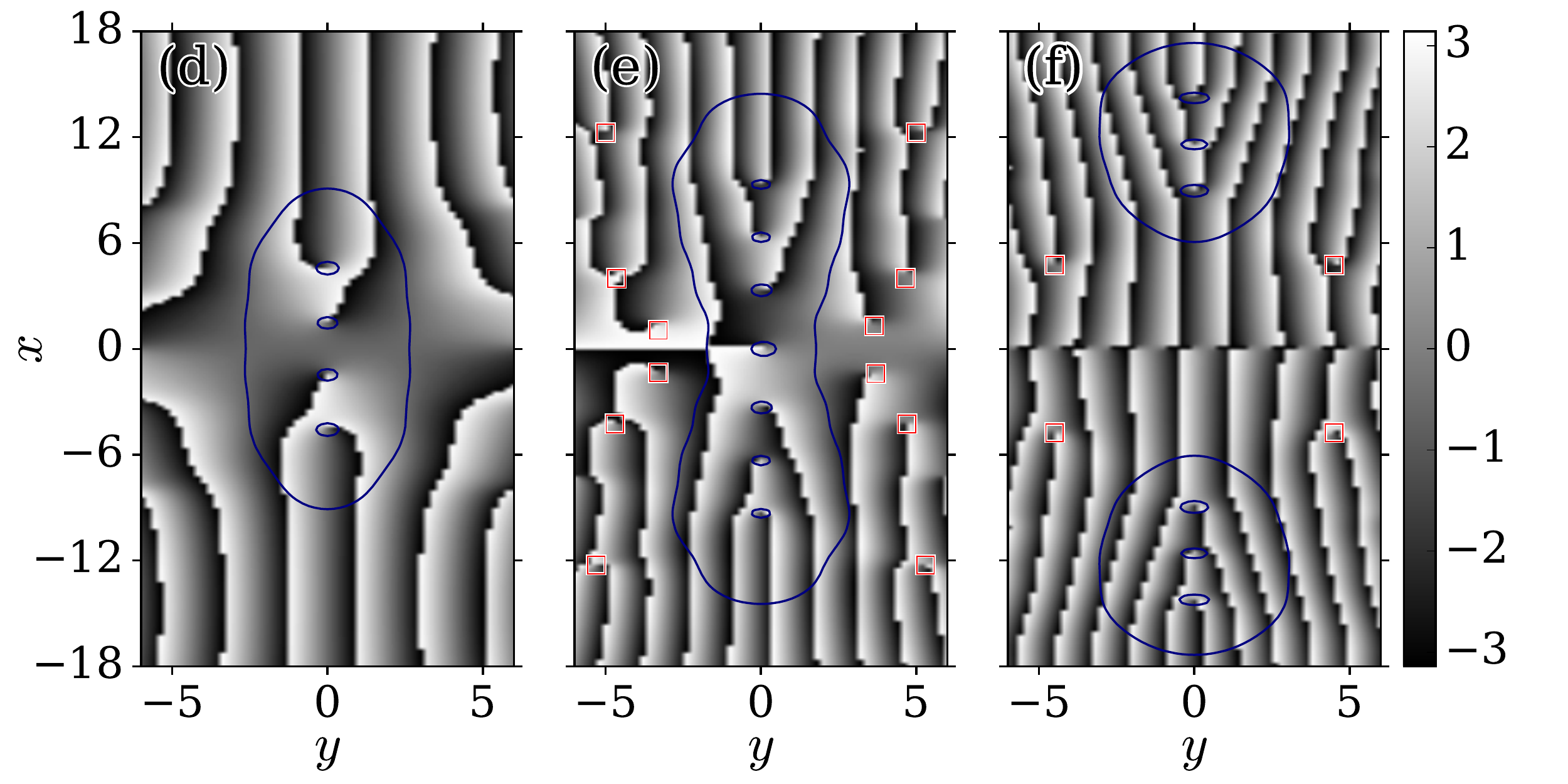} \end{center} \caption{(Color online) Vortex 
lattices in rotating dipolar BECs trapped in asymmetric trap with $\gamma = 
0.5$, $\lambda=10$, $D=38$ and rotation frequency: (a) $\Omega =0.425$, (b) $\Omega =0.6$, (c) $\Omega 
=0.7$. The contour levels are taken at $\vert\phi({x,y,z})\vert ^2 = 0.01$, $0.005$ and $0.002$, 
respectively. The corresponding phase of the condensate density $\vert\phi({x,y,0})\vert ^2$ is shown 
in d-f. Blue lines are density contour lines. Blue circles show the position 
of vortices, red squares show the center of hidden vortices.
} \label{fig11} 
\end{figure} 
The condensate strongly elongates along the $x$-direction at 
$\gamma \leq 0.5$. The centrifugal force also elongates the condensate along $x$-direction. We have 
shown four linearly arranged vortices in figure~\ref{fig11}(a) for $\gamma = 0.5$ and $\Omega=0.425$. 
In the figures~\ref{fig11} and \ref{fig12}, we observed linear and zig-zag arrangements of vortices with 
respect to the strength of eccentricity of the trap, respectively. Such linear and zig-zag vortex 
configurations have been observed in conventional quasi two-dimensional BEC~\cite{aniso-bec}.
Further increase in $\Omega$ splits the condensate at a critical rotation frequency 
($\Omega_{sp}$). In addition to the repulsive dipolar force between condensate atoms, the rotation 
provides supplementary elongation at $\Omega$ $\geq$ $\Omega_{sp}$. Splitting is a combined effect of rotation and dipolar
forces. As rotation increases, the condensate creates extra vortices and also stretches. In the
limiting case, it remains a vortex in the center and further stretching breaks the condensate in
two parts. With further increase of rotation, no more vortices are generated since extra angular 
momentum can now be accommodated by orbital angular momentum. Stretching is enhanced depending on the trap geometries. 
We have observed only two vortices as shown in figure~\ref{fig12}(a) at $\gamma = 0.7$ for the rotation frequency $\Omega=0.425$.  
\begin{figure}[htpb]
\begin{center} 
\includegraphics[width=0.25\columnwidth]{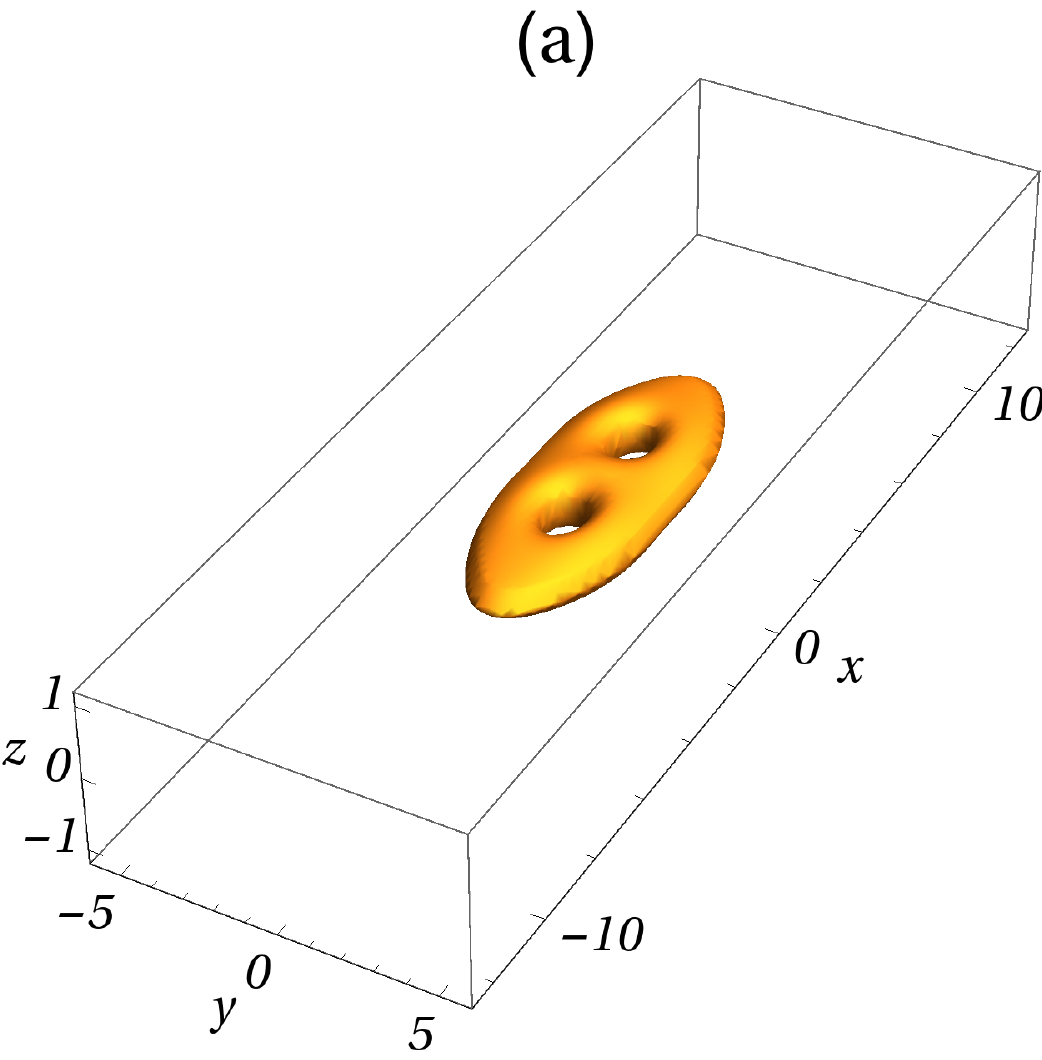} 
\includegraphics[width=0.25\columnwidth]{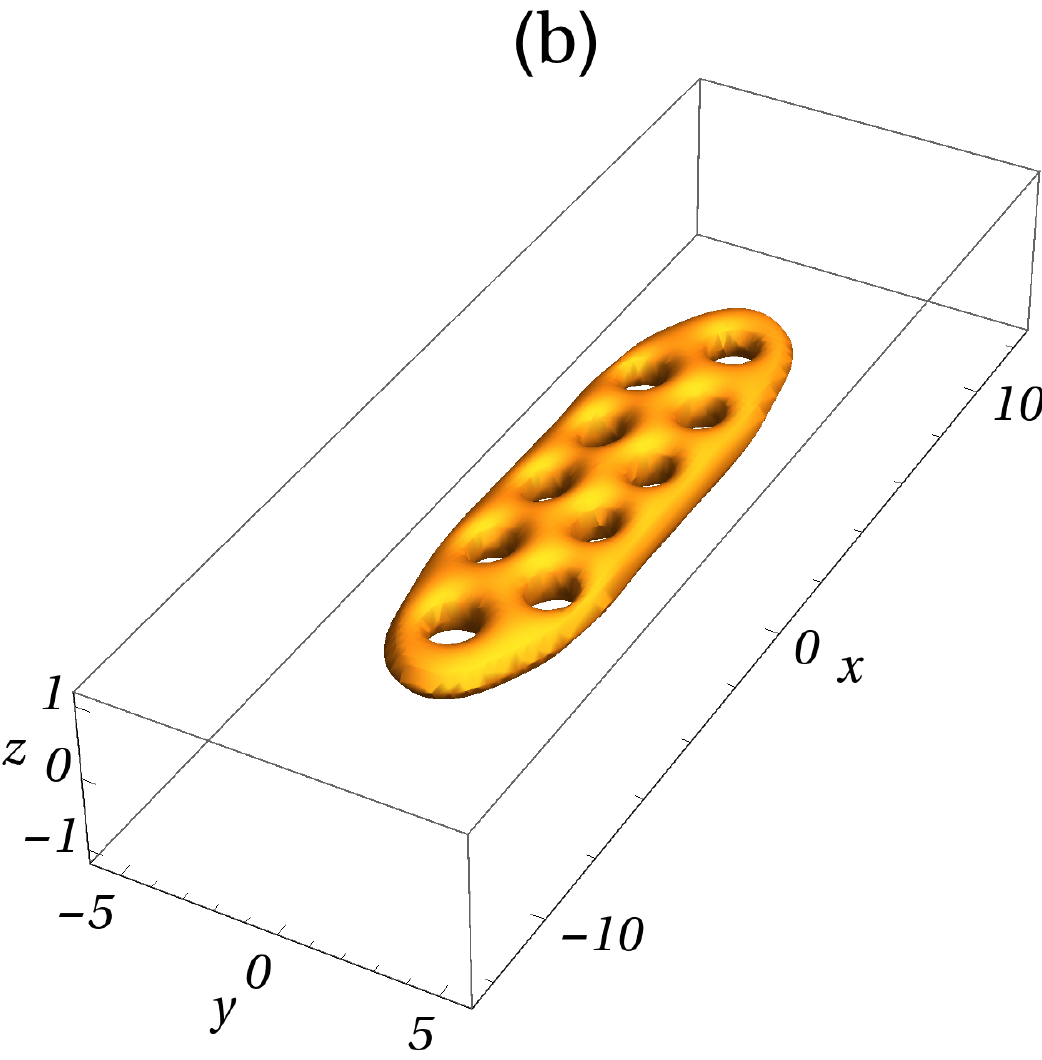} 
\includegraphics[width=0.25\columnwidth]{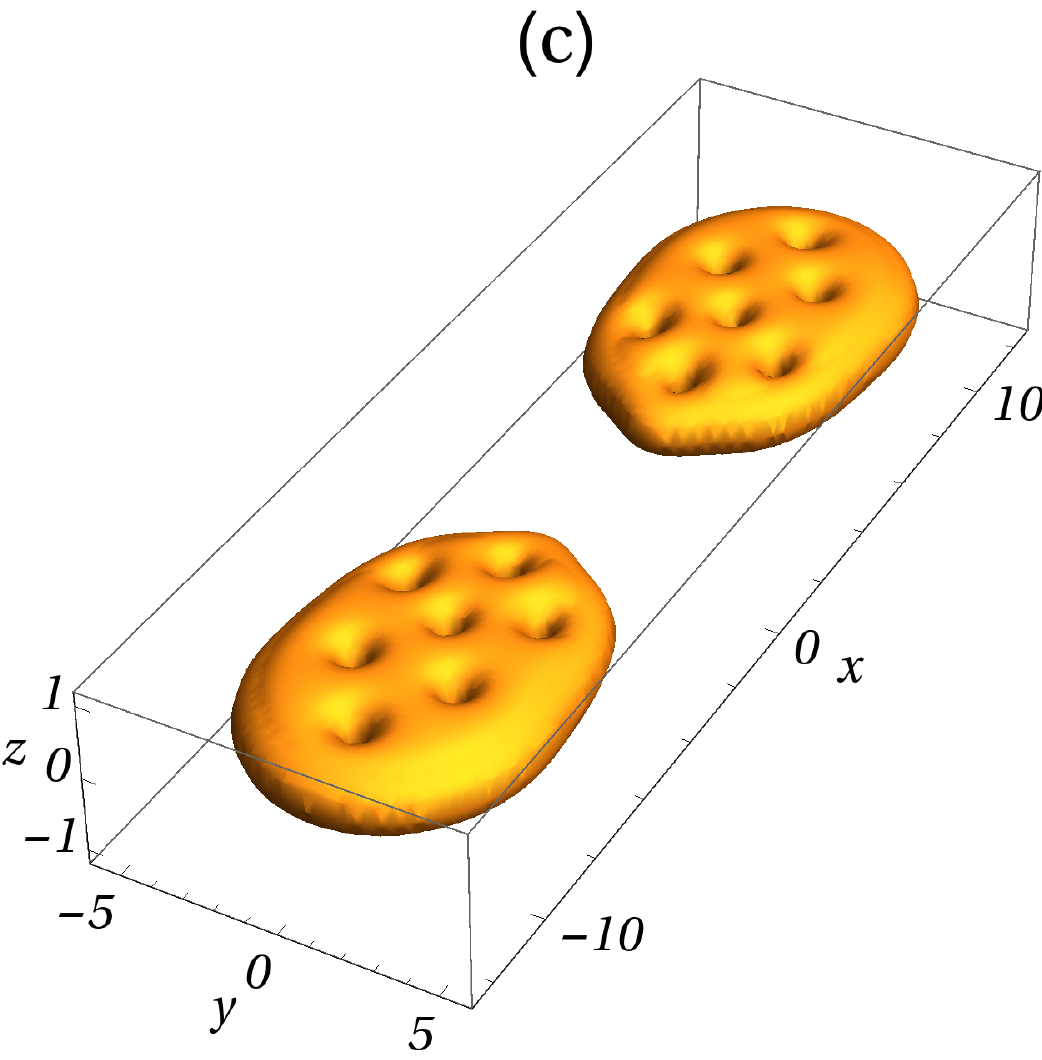} 
\includegraphics[width=0.65\columnwidth]{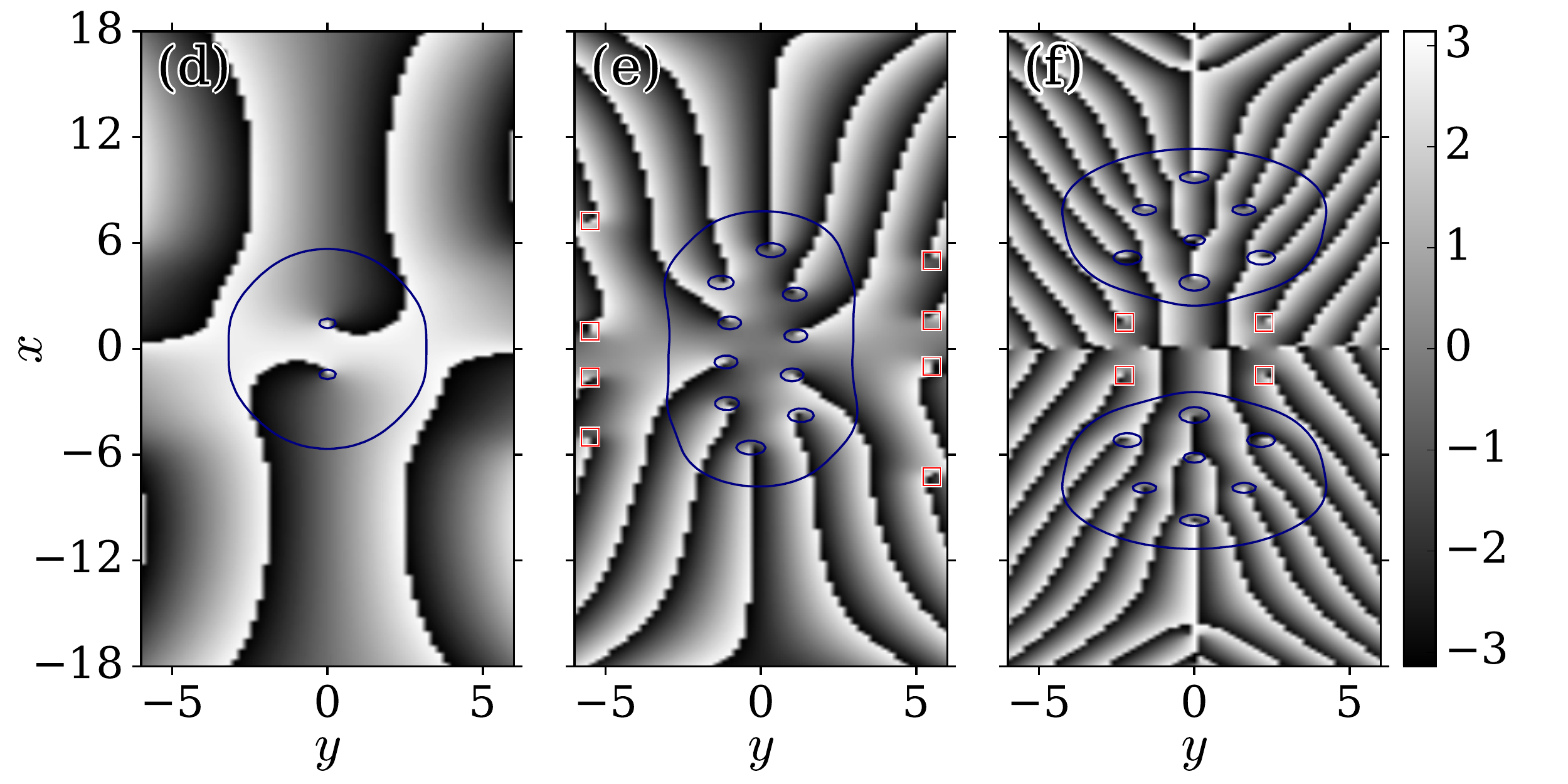} \end{center} 
\caption{(Color online) Vortex 
lattices in rotating dipolar BECs trapped in asymmetric trap with $\gamma= 
0.7$, $\lambda=10$, $D=38$ and rotation frequency: (a) $\Omega = 0.425$, (b) $\Omega = 0.7$ , (c) 
$\Omega = 0.91$. The contour levels are taken at $\vert\phi({x,y,z})\vert ^2 = 0.02$, $0.012$ and 
$0.002$, respectively. The corresponding phase of the condensate density $\vert\phi({x,y,0})\vert ^2$ 
is shown in (d-f). Blue lines are density contour lines. Blue circles show the position 
of vortices, red squares show the center of hidden vortices.}
\label{fig12}
\end{figure}%
This is exactly half the number of vortices when compared to that with $\gamma = 0.5$ for the same 
rotation frequency. This evidences the dependence of the number of vortices on the
eccentricity of the trap. At sufficiently high $\Omega_{sp}$, the condensate splits into two parts 
as shown in figure~\ref{fig12}(c). Here the splitting occurs at $\Omega_{sp}=0.91$, which indicates 
that the trap with large $\gamma$ requires substantially larger rotation frequency than the condensate 
with smaller $\gamma$. The phase of the vortices, marked with circles, is shown in 
figures~\ref{fig11}(d)-(f) and \ref{fig12}(d)-(f). We also marked few more vortices as red 
squares.  These vortices are called hidden (or ghost) vortices since they cannot be observed in 3D density 
distribution. Due to hidden vortices, there is a phase defect distributed along the $y$-axis in the split 
condensate. A similar phase distribution is observed in rotating BEC in a 
double-well potential~\cite{dwell}. The splitting with vortices are also observed in normal BEC, where it is triggered by random 
noise or a complex frequency to the stationary vortices~\cite{split-vort}. However, splitting is spontaneous in rotating dipolar 
BEC.
\begin{figure}[!ht]
\begin{center}
\includegraphics[width=0.65\columnwidth]{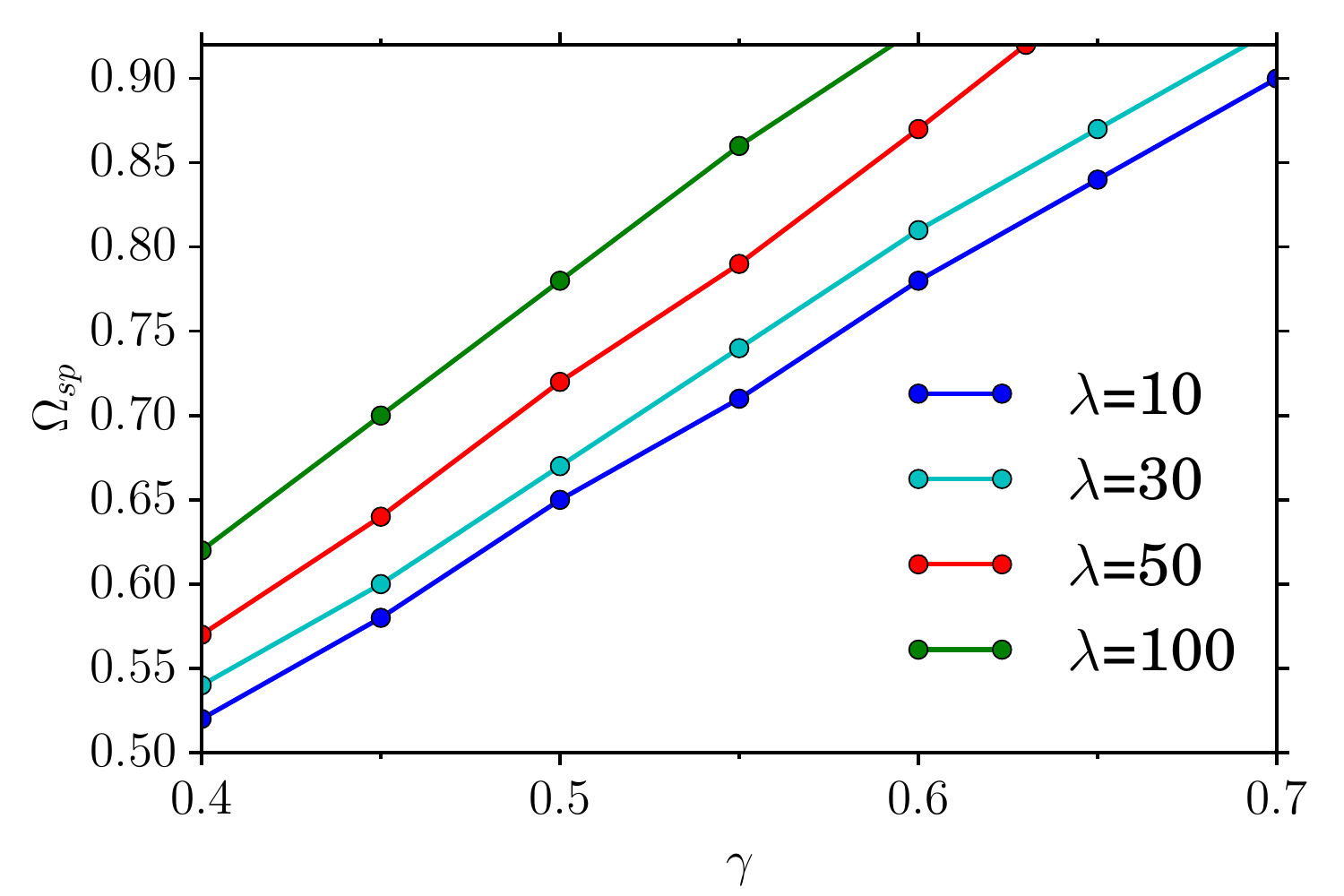}
\end{center}
\caption{ Critical rotation frequency ($\Omega_{sp}$) for the splitting of the condensate with
 respect to anisotropy ($\gamma$) of the trap along $x$-direction for the parameters $\nu=1$, $D=38$ and for different 
trap aspect ratios $\lambda$ $=10$, $30$, $50$ and $100$.}
\label{fig13}
\end{figure}%

In figure~\ref{fig13}, we show $\Omega_{sp}$ at different trap aspect 
ratios, $\lambda$ $=10$, $30$, 
$50$ and $100$ with respect to the eccentricity of the trap $\gamma$. 
When the anisotropy is fixed as $\gamma = 0.5$ and $\lambda = 10$ then the 
splitting is observed at $\Omega_{sp}=0.65$.  One can observe that $\Omega_{sp}$ increases with increasing trap aspect ratio. 
For instance, when 
$\lambda = 50$ and $100$ the corresponding required $\Omega_{sp}$ are $0.72$ and $0.78$, respectively. The strong axial trap preserves 
the condensate from splitting, and one needs to apply stronger $\Omega$ to observe it. On the other hand, if we increase 
the anisotropy to $\gamma=0.7$, the $\Omega_{sp}$ also increases to 
$0.9$ for $\lambda=10$. The condensate does not split in strong enough axial traps  
as for $\gamma=0.7$ and $\lambda$ = $50$ and $100$. 
We have also carried out the calculations with an effective 2D GP equation in a rotating frame and observed a similar 
splitting with same vortex distribution for the corresponding parameters used in 3D calculations, confirming our results.
\section{Summary and Conclusion}

\label{sec:con}

In summary, we have studied the vortex lattice structures in purely dipolar Bose-Einstein condensate of $^{52}$Cr 
atoms by considering the full three-dimensional Gross-Pitaevskii equation. We have identified the stability regimes 
for non-vortex as well as vortex states in purely dipolar BECs with respect to both trap aspect ratio and dipole-dipole 
interaction strength. The stability of vortex lattice structures are confirmed by real time evolution. We have 
shown stationary vortex lattice structures for different trap aspect ratios within the stability regime. Also, 
distortions in the lattices at high rotation frequencies have been observed. %

We estimated the number of vortices using Thomas-Fermi approximation and compared with those from numerical simulations. 
Further, we have analysed the spatial distribution of vortices in the fully anisotropic trap and 
observed linear and zig-zag arrangement of vortices. We noticed the breaking up of the condensates into two parts with an equal number of 
vortices. The splitting occurs due to a combination of 
centrifugal force due to rotation and repulsive dipolar interaction. Increasing rotation frequency 
creates extra vortices and stretches. Close to $\Omega_{sp}$ it appears a 
vortex at the center and further stretching breaks the condensate in two parts. Critical rotation frequency for splitting depends 
on the trap aspect ratio and dipolar parameters.

The predicted phase diagram of the stable vortex state will be useful to demonstrate the parameters such as dipolar 
strengths and trap aspect ratios for making experimental and theoretical studies on rotating dipolar quantum gases. 
Further, the stability regime is relevant to investigate rapidly rotating dipolar BECs in the lowest 
Landau level. We found that dipolar BECs under rotation can produce square vortex lattices. This was found with
two-component BECs~\cite{2c-square} and recently in dipolar Fermi gas \cite{Kohn-sham}.
Observing the breaking in the rotating dipolar BECs in a fully anisotropic trap will be a new
experimental exploration.

\ack
RKK acknowledges the financial support from FAPESP of Brazil (Contract number 2014/01668-8). TS acknowledges financial support 
from University Grants Commission, India in the form UGC-RFSMS fellowship. The work of PM forms a part of Science \& Engineering 
Research Board, Department of Science \& Technology, Govt. of India sponsored research project (SERB Ref. No. EMR/2014/000644). AG 
and HF acknowledges CAPES, CNPq and FAPESP of Brazil.

\end{document}